\documentclass[twocolumn]{aastex62}

\usepackage{amsmath}
\usepackage{mathtools}
\usepackage{amsmath}
\usepackage{graphicx}
\usepackage{hyperref}

\begin{document}

\title{On the GeV emission of the type I BdHN GRB 130427A}

\author{R.~Ruffini}
\affiliation{ICRA, Dipartimento di Fisica, Sapienza Universit\`a di Roma, P.le Aldo Moro 5, 00185 Rome, Italy}
\affiliation{ICRANet, P.zza della Repubblica 10, 65122 Pescara, Italy}
\affiliation{Universit\'e de Nice Sophia Antipolis, CEDEX 2, Grand Ch\^{a}teau Parc Valrose, Nice, France}
\affiliation{ICRANet-Rio, Centro Brasileiro de Pesquisas F\'isicas, Rua Dr. Xavier Sigaud 150, 22290--180 Rio de Janeiro, Brazil}
\affiliation{INAF, Viale del Parco Mellini 84, 00136 Rome, Italy}

\author{R.~Moradi},\email{rahim.moradi@icranet.org}
\affiliation{ICRA, Dipartimento di Fisica, Sapienza Universit\`a di Roma, P.le Aldo Moro 5, 00185 Rome, Italy}
\affiliation{ICRANet, P.zza della Repubblica 10, 65122 Pescara, Italy}

\author{J.~A.~Rueda}\email{jorge.rueda@icra.it}
\affiliation{ICRA, Dipartimento di Fisica, Sapienza Universit\`a di Roma, P.le Aldo Moro 5, 00185 Rome, Italy}
\affiliation{ICRANet, P.zza della Repubblica 10, 65122 Pescara, Italy}
\affiliation{ICRANet-Rio, Centro Brasileiro de Pesquisas F\'isicas, Rua Dr. Xavier Sigaud 150, 22290--180 Rio de Janeiro, Brazil}
\affiliation{INAF, Istituto di Astrofisica e Planetologia Spaziali, Via Fosso del Cavaliere 100, 00133 Rome, Italy.}

\author{L.~Becerra}
\affiliation{Escuela de F\'isica, Universidad Industrial de Santander, A.A.678, Bucaramanga, 680002, Colombia}

\author{C.~L.~Bianco}
\affiliation{ICRA, Dipartimento di Fisica, Sapienza Universit\`a di Roma, P.le Aldo Moro 5, 00185 Rome, Italy}
\affiliation{ICRANet, P.zza della Repubblica 10, 65122 Pescara, Italy}
\affiliation{INAF, Istituto di Astrofisica e Planetologia Spaziali, Via Fosso del Cavaliere 100, 00133 Rome, Italy.}

\author{C.~Cherubini}\author{S.~Filippi}
\affiliation{ICRANet, P.zza della Repubblica 10, 65122 Pescara, Italy}
\affiliation{ICRA and Department of Engineering, University Campus Bio-Medico of Rome, Via Alvaro del Portillo 21, 00128 Rome, Italy}

\author{Y.~C.~Chen}
\affiliation{ICRA, Dipartimento di Fisica, Sapienza Universit\`a di Roma, P.le Aldo Moro 5, 00185 Rome, Italy}
\affiliation{ICRANet, P.zza della Repubblica 10, 65122 Pescara, Italy}

\author{M.~Karlica}
\affiliation{ICRA, Dipartimento di Fisica, Sapienza Universit\`a di Roma, P.le Aldo Moro 5, 00185 Rome, Italy}
\affiliation{ICRANet, P.zza della Repubblica 10, 65122 Pescara, Italy}
\affiliation{Universit\'e de Nice Sophia Antipolis, CEDEX 2, Grand Ch\^{a}teau Parc Valrose, Nice, France}

\author{N.~Sahakyan}
\affiliation{ICRANet, P.zza della Repubblica 10, 65122 Pescara, Italy}
\affiliation{ICRANet-Armenia, Marshall Baghramian Avenue 24a, Yerevan 0019, Armenia}

\author{Y.~Wang}
\affiliation{ICRA, Dipartimento di Fisica, Sapienza Universit\`a di Roma, P.le Aldo Moro 5, 00185 Rome, Italy}
\affiliation{ICRANet, P.zza della Repubblica 10, 65122 Pescara, Italy}

\author{S.~S.~Xue}
\affiliation{ICRA, Dipartimento di Fisica, Sapienza Universit\`a di Roma, P.le Aldo Moro 5, 00185 Rome, Italy}
\affiliation{ICRANet, P.zza della Repubblica 10, 65122 Pescara, Italy}


\begin{abstract}
We propose that the \textit{inner engine} of a type I binary-driven hypernova (BdHN) is composed of {a Kerr black hole (BH) in a non-stationary state, embedded in a uniform magnetic field $B_0$ aligned with the BH rotation axis, and surrounded by an ionized plasma of extremely low density of $10^{-14}$~g~cm$^{-3}$. 
}
Using GRB 130427A as a prototype we show that this \textit{inner engine} acts in a sequence of \textit{elementary impulses}. Electrons are accelerated to ultra-relativistic energy near the BH horizon and, propagating along the polar axis, ${\theta =0}$, they can reach energies of $\sim 10^{18}$~eV, and partially contribute to ultra-high energy cosmic rays (UHECRs). When propagating with ${\theta \neq 0}$ {through the magnetic field $B_0$ they} give origin by synchrotron emission to GeV and TeV radiation. {The mass of BH, $M=2.3 M_\odot$, its spin, $\alpha = 0.47$, and the value of magnetic field $B_0= 3.48 \times 10^{10}$~G, are determined self-consistently in order to fulfill the energetic and the transparency requirement.} {The} repetition time {of each elementary impulse  of energy ${\cal E} \sim 10^{37}$~erg, is $\sim 10^{-14}$~s at the beginning of the process, then} slowly increasing with time evolution. In principle, this ``\textit{inner engine}'' can operate in a GRB for thousands of years.  By scaling the BH mass and the magnetic field the same ``\textit{inner engine}'' can describe  active galactic nuclei (AGN).
\end{abstract}



\section{Introduction}\label{sec:1}

\begin{figure*}
    \centering
    \includegraphics[width=\hsize,clip]{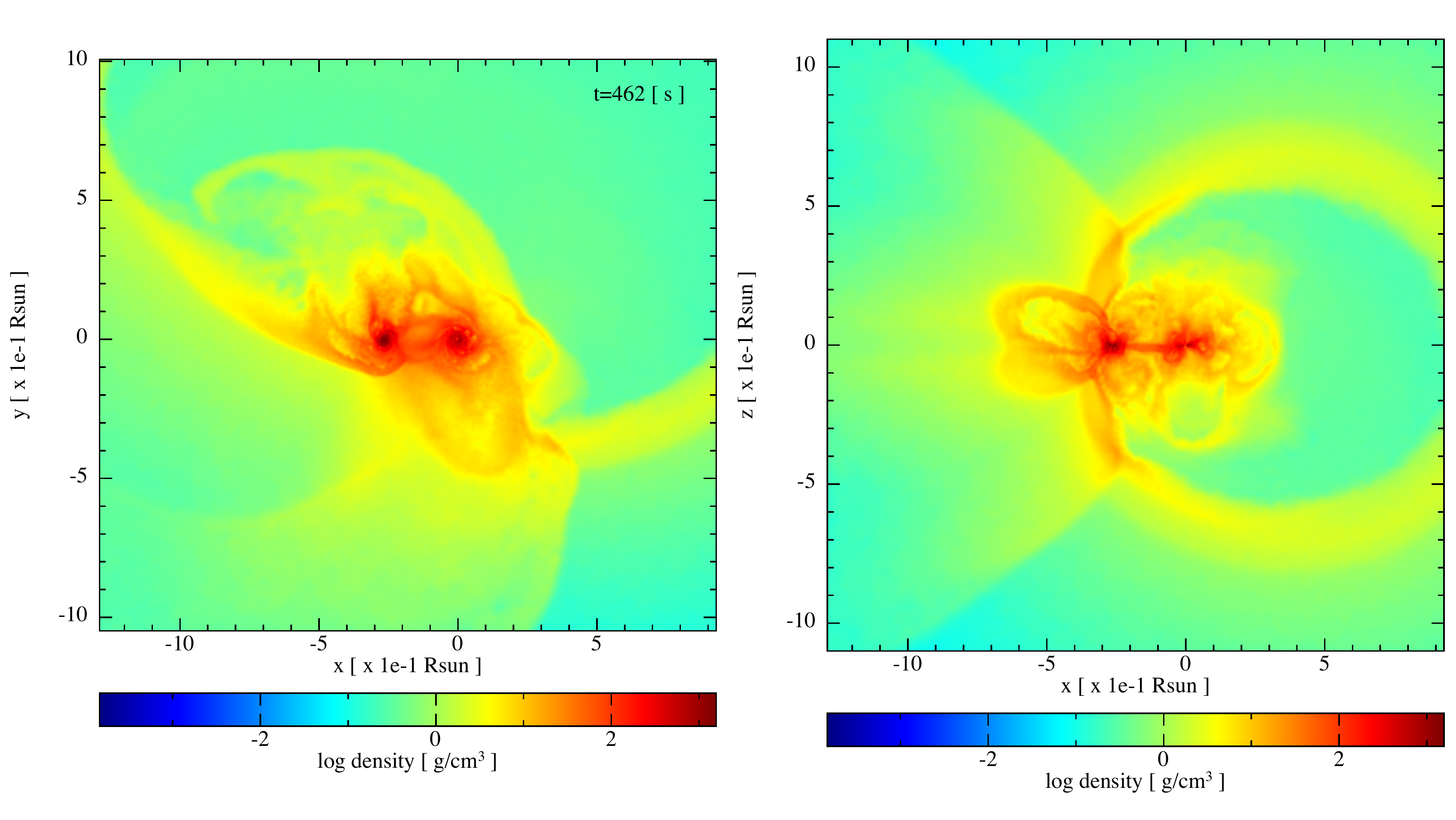}
    \caption{Selected SPH simulation from \citet{2019ApJ...871...14B} of the exploding CO$_{\rm core}$ as SN in the presence of a companion NS: Model `25m1p07e' with $P_{\rm orb}\approx 5$~min. The CO$_{\rm core}$ is taken from the $25~M_\odot$ zero-age main-sequence (ZAMS) progenitor, so it has a mass $M_{\rm CO}=6.85~M_\odot$. The mass of the NS companion is $M_{\rm NS}=2~M_\odot$. The plots show the surface density on the equatorial binary plane (left panel) and on a plane orthogonal to it (right panel) at the time in which the NS companion reaches the critical mass and collapses to a BH, $t=462$~s from the SN shock breakout ($t=0$ of our simulation). The coordinate system has been rotated and translated in such a way that the NS companion is at the origin and the $\nu$NS is along the $-x$ axis.}
    \label{fig:BdHN}
\end{figure*}

Nine subclasses of gamma-ray bursts (GRBs) with binary progenitors  have been recently introduced in \citet{2016ApJ...832..136R,2018ApJ...859...30R,2018JCAP...10..006R1,2019ApJ...874...39W}.  One of the best prototypes of the long GRBs emitting $0.1$--$100$~GeV radiation is  GRB 130427A \citep{2015ApJ...798...10R}. It belongs to a special subclass of GRBs originating from a tight binary system, of orbital period $\sim 5$~min, composed of a carbon-oxygen core (CO$_{\rm core}$) undergoing a supernova (SN) event, {in  presence of a} neutron star (NS) {companion. The SN, as usual, gives rise to a new NS ($\nu$NS)}. {For binary periods $\lesssim 5$~min}  the hypercritical accretion of the SN ejecta onto the companion NS leads it to exceed the critical mass for gravitational collapse and form a Kerr black hole (BH). We have called these systems binary-driven hypernovae of type I (BdHNe I) {with $E_{\rm iso}>10^{52}$~erg as opposed to BdHNe II with binary periods $\gtrsim 5$~min and $E_{\rm iso}<10^{52}$~erg when the NS critical mass is not exceeded} \citep{2019ApJ...874...39W}. Figure~\ref{fig:BdHN} shows the ejecta density distribution of a BdHN I on the binary equatorial plane (left panel) and in a plane orthogonal to it (right panel), at the moment of gravitational collapse of the NS companion, namely at the moment of BH formation. These plots are the result of three-dimensional, numerical smoothed-particle-hydrodynamic (SPH) simulations of BdHNe recently described in \citet{2019ApJ...871...14B}. 

In the specific case of GRB 130427A this BdHN I is seen from ``the top'' with the viewing angle in a plane orthogonal to the plane of the orbit of the binary progenitor. This allows us to follow all the details of the high energy activities around the BH. This includes: {a) first appearance of the supernova (the \textit{SN-rise})}, b) the observation of the ultrarelativistic prompt emission (UPE) following the BH formation \citep{ 2019arXiv190404162R} , c) the feedback of the SN ejecta {accreting} onto the $\nu$NS leading to the X-ray afterglow \citep{2018ApJ...869..101R}, d) the ultra-high energy process extracting the rotational energy of the BH, { reducing its mass and spin,} and generating the GeV and TeV radiation {are presented in this article.}

Soon after the BH formation, approximately $10^{57}$ baryons, which include the ones composing the NS companion, are enclosed in the BH horizon beyond any possible measurable effect apart from the total mass and spin of the BH. 

A cavity of approximately $10^{11}$~cm is formed around the BH with a finite density of $10^{-6}$~g~cm$^{-3}$, see \citep{2018ApJ...852..120B, 2019ApJ...871...14B}. The evolution of such a cavity following the GRB explosion and its overtones inside the cavity has been addressed in the joint article \citep{2019ApJ...883..191R}, finally reaching a density of $10^{-14}$~g~cm$^{-3}$ inside the cavity.

The Kerr BH formation occurs in such a cavity in presence of an external uniform magnetic field aligned with the BH rotation axis {which has been estimated in this paper to be $B_0\sim 10^{10}$~G,}. This \textit{composite system} is the ``\textit{inner engine}'' of the GRB. For quantitative estimates, we consider it as mathematically described by a non-stationary {Papapetrou solution} \citep{1966AIHPA...4...83P,Wald:1974np,2019arXiv190511339R}.

As we will quantify later in this article a sufficient amount of low density ionized matter will be needed in the cavity in order to feed this \textit{inner engine}.

{In this article we assume that the magnetic field and the BH spin are parallel. In that case the induced electric field 
is such that electrons(protons) along and near the rotation axis in the surrounding ionized circumburst medium are repelled(attracted) by the BH. The behavior is vice versa in the anti-parallel case. 
} As pointed out by \citet{2013CQGra..30l5008G}, the stability of Wald-like solutions is guaranteed only if the uniform field is confined to a radius smaller than the Melvin radius, 
\begin{equation}\label{eq:RM}
R_M\sim 2/B_0,    
\end{equation}
which imposes an upper limit for this geometry of about {$10^{15}$~cm} in our present case\footnote{The conversion factor from CGS to geometric units for the magnetic field is: $\sqrt{G}/c^2 \approx 2.86\times 10^{-25}$, where $G$ and $c$ are the gravitational constant and speed of light in CGS units, respectively. Therefore a magnetic field on the order of {$10^{10}$~G} in geometric units is $\approx 2\times 10^{-25}\times 10^{10}\approx 2 \times 10^{-15}$~cm$^{-1}$ which leads to the Melvin radius of {$R=2/B_0 \approx 10^{15}$~cm}.}. {We are going to show in this article} that the particle acceleration occurs near the BH horizon within distance {of approximately $10^{5}$~cm}, much smaller than $R_M$.

We recall the definition of the critical electric and magnetic fields, $E_c$ and {$B_{c}$}, i.e. 
\begin{equation}\label{eq:Ec}
 E_{c}=\frac{m_e^2 c^3}{e\hbar},\quad {B_{c}=\frac{m_e^2 c^2}{e\hbar}={4.4\times 10^{13}~\rm G}}
\end{equation}
where $m_e$ and $e$ are the electron mass and charge, respectively.

{
Particular attention is devoted to identify the regimes in which the electric and magnetic fields are undercritcal or overcritical and,correspondingly,  study the pair creation process and the associated absorption or transparency conditions for the GeV emission. In this article we will focus on the undercritical regime in GRB 130427A
}

{
Our main goal is to develop an ``\textit{inner engine}'' model consistent with the transparency condition of the GeV and high-energy emissions from GRB 130427A. The model makes use of:
}

1) the rotational energy of the BH as its energy source;

2) the acceleration {and radiation processes of ultrarelativistic electrons near the horizon of the BH and in presence of the uniform magnetic field $B_0$, determined by using the electrodynamical properties of the Wald solution and}

3) the determination of the highly anisotropic  GeV, TeV and UHECR emission by the synchrotron radiation, as a function of the injection angle of the ultra relativistic electrons.

As a byproduct, we show:

1) {that the high-energy emission of GRB 130427A, far from being emitted continuously, actually occurs in a repetitive sequence of discrete, ``quantized'', ``elementary impulsive events'' or, for short, ``quanta'';}

2) {each ``quantum'' carries an energy of the order of $10^{37}$~erg;}

3) {each ``quantum'' is repetitively emitted with a repetition time $\sim 10^{-14}$~s}.

{The three fundamental parameters of the model, i.e. the Kerr BH mass, $M$, spin parameter, $\alpha = c\,J/(G\,M^2)$ where $J$ is the BH angular momentum, and the magnetic field $B_0$, are determined as follows:}

1) {The magnetic field $B_0$ is obtained by imposing the transparency condition of the GeV luminosity and as well the coincidence between the theoretical predicted repetition time of the ``quanta'' and the time scale of first impulsive event. }

2) {The BH mass $M$ and spin parameter $\alpha$, as well as their their temporal evolution, are determined by obtaining the  GeV luminosity via the extractable energy of the BH.}

3) {In each one of these \textit{elementary impulsive events} we can estimate the depletion of the rotational energy of the Kerr BH, consequently we can estimate that the high-energy emission process can indeed last for thousands of years.}

{The article is organized as follows. In section~\ref{sec:countrate} we recall the count rate and light-curves of Fermi-GBM and Fermi-LAT for GRB 130427A.}
{In section ~\ref{sec:2} the basic equations for determining the extraction of rotational energy from a Kerr BH in order to explain the GeV energetic are expressed in terms of the BH mass and spin}. In section~\ref{innerengine} the electrodynamics of the ``inner engine'' is presented. {In section~\ref{sec:4} the basic equations governing the synchrotron radiation in {the magnetic field $B_0$}, the first elementary event and the limit on the magnetic field in order to ground the transparency of the GeV radiation are established.}
{In section~\ref{sec:massspin} we determine the mass and spin of the BH in order to fulfill the GeV energetic and we address the decrease of the mass and spin of the BH as a function of the extracted rotational energy.} 
{In section~\ref{sec:power} the synchrotron radiation power and the need of a low density ionized plasma in order to explain the number of needed electrons to feed the system is presented.}
In section~\ref{sec:6} the sequence of {``\textit{quanta}''} and their repetition time are indicated. We also outline the mounting evidence that this system, here developed and applied for GRB 130427A, may well be extended to the much more massive BHs of $10^9\,M_\odot$ in AGN such as M87.

{\section{Count rate and light-curves of Fermi-GBM and Fermi-LAT}\label{sec:countrate}}

\begin{figure*}
\centering
[a]\includegraphics[width=0.75\hsize,clip]{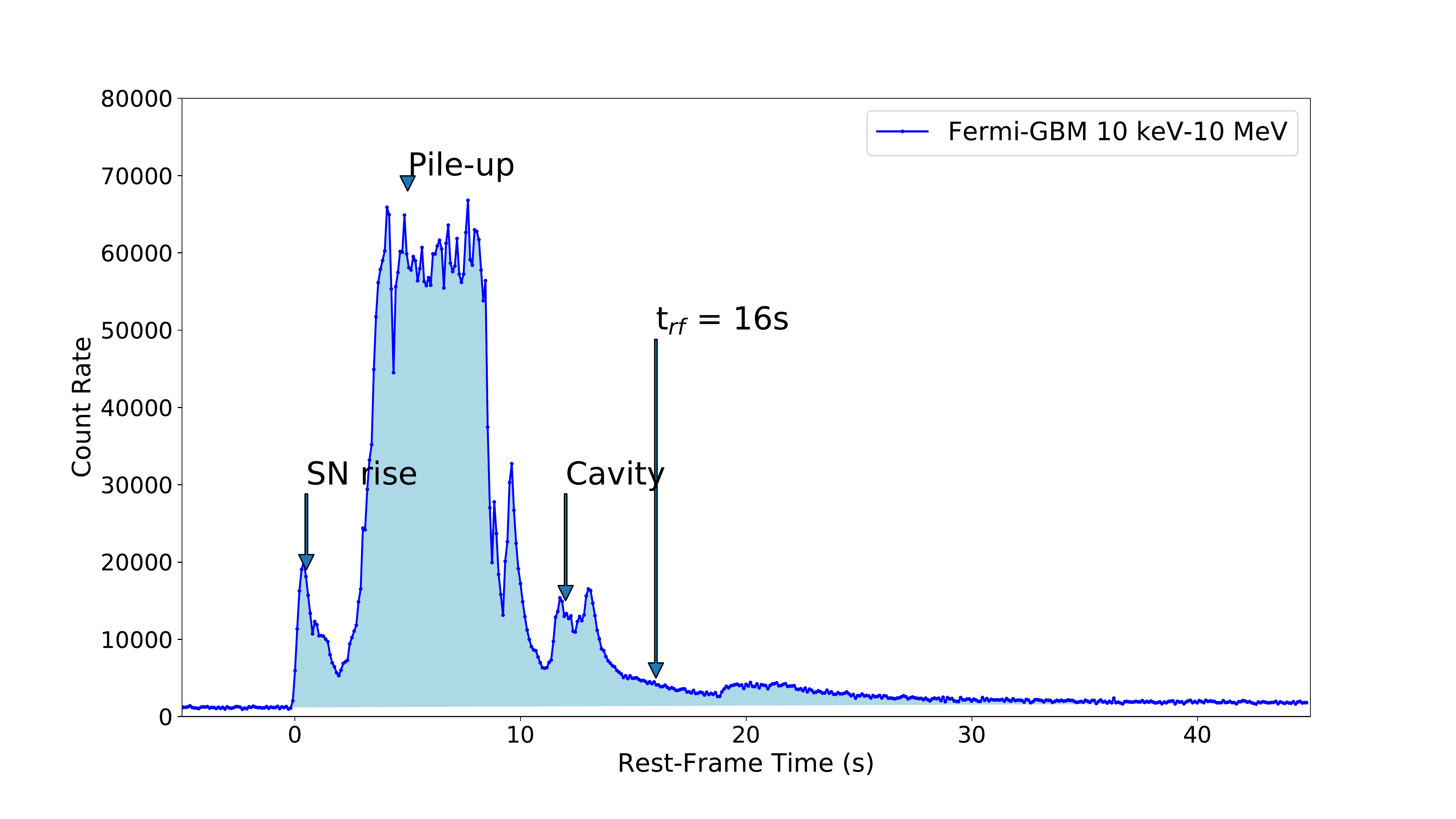}
[b]\includegraphics[width=0.75\hsize,clip]{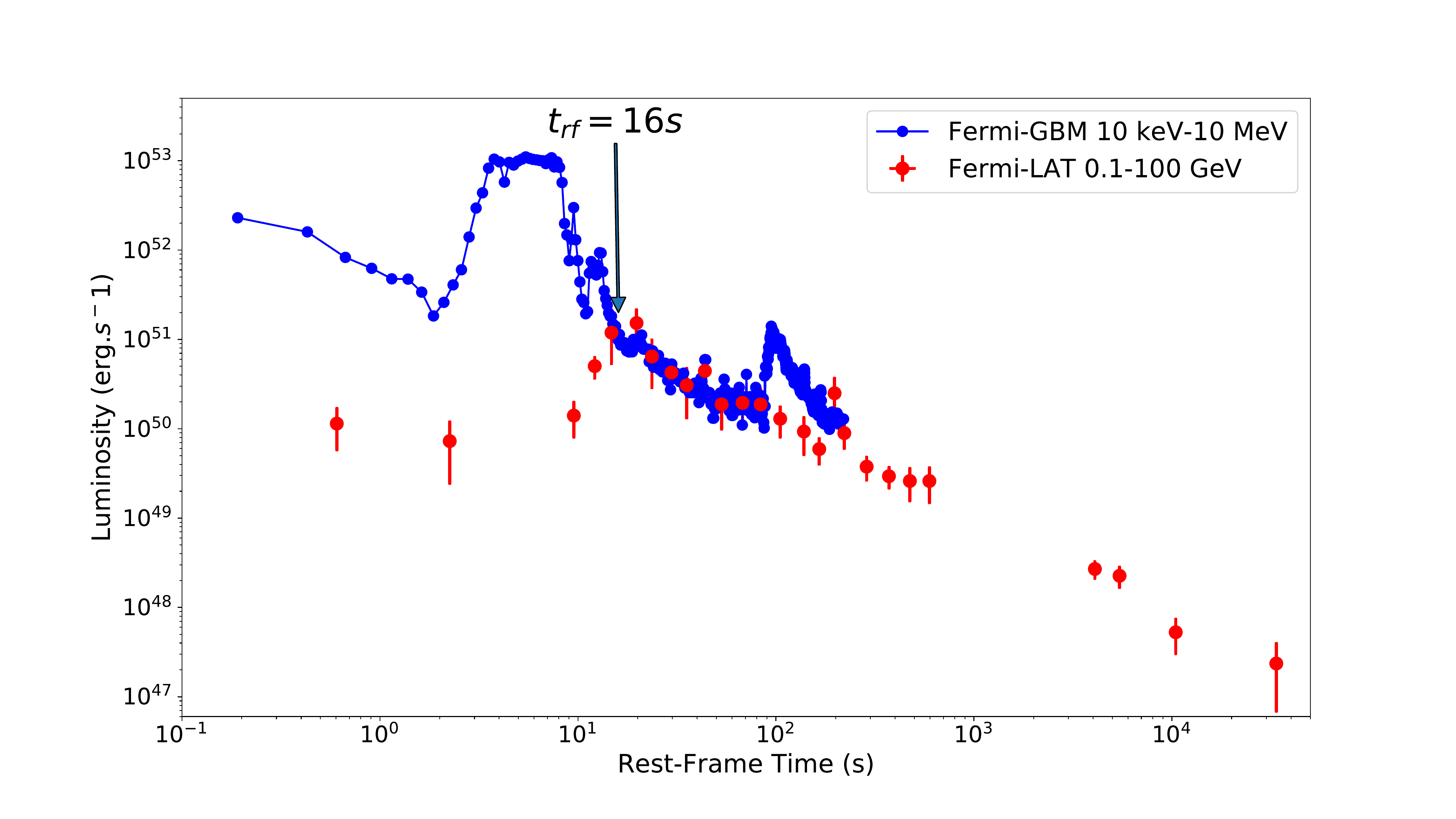}
[c]\includegraphics[width=0.75\hsize,clip]{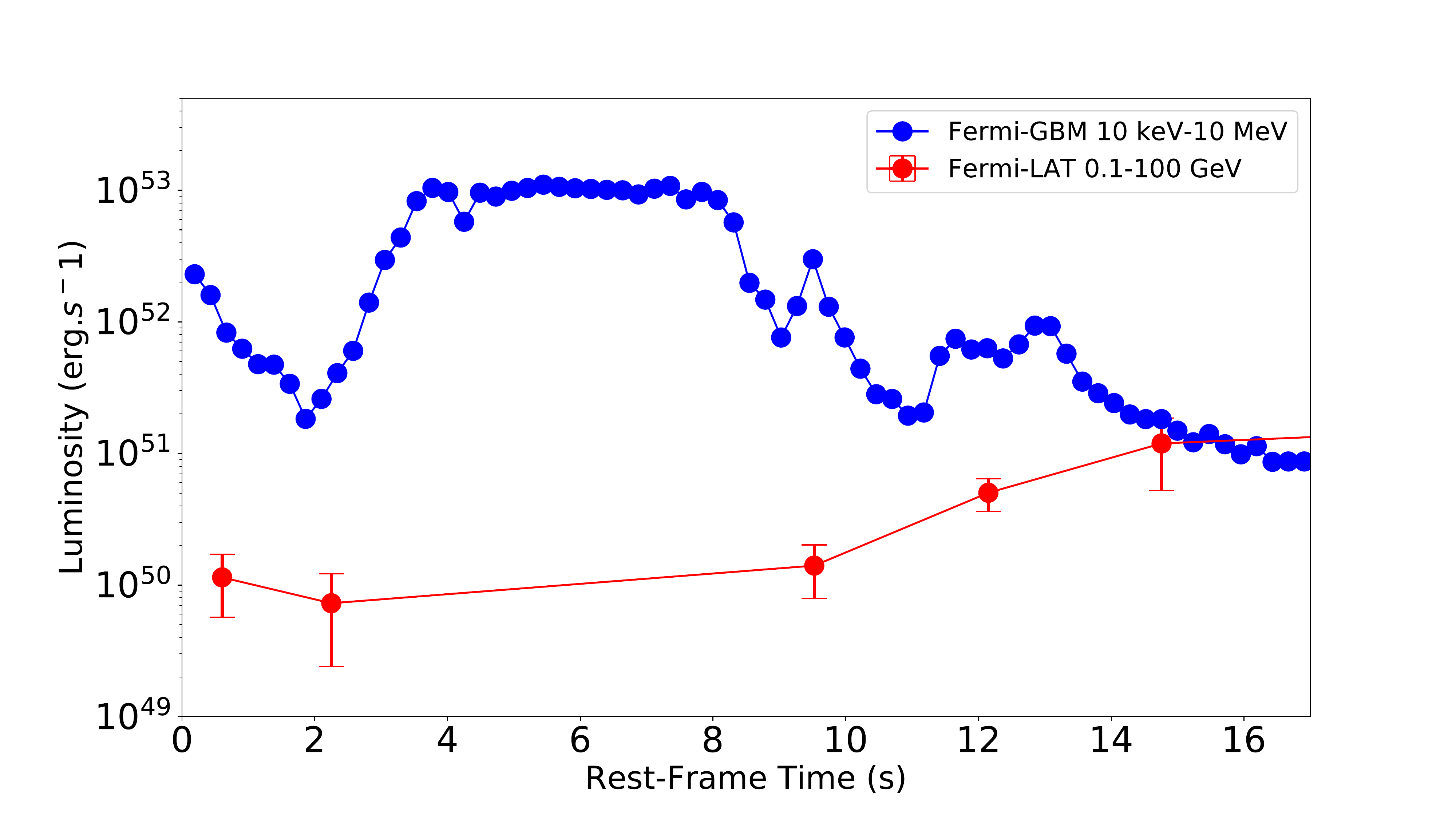}
\caption{a: {The Fermi-GBM} count rate of GRB 130427A. In rest-frame time  {interval} $[T_0 + 3.4$~s,$T_0 + 8.6$~s], the GRB is affected by pile-up. {\textbf{b:} The luminosity of GRB 130427A in the Fermi energy range. \textbf{c:} The anti-correlation between the flux (luminosity) received by Fermi-GBM and Fermi-LAT in the time interval [1s, 16s], indicates that the primary photons in the GeV energy range are converted to the MeV photons due to the high opacity, details in \citep{ 2015ApJ...798...10R}.}}
\label{prompt}
\end{figure*} 

As detailed in \citet{2013GCN..14455...1L, 2013GCN.14473....1V, 2013GCN..14478...1X, 2013GCN..14491...1F, 2015ApJ...798...10R} GRB 130427A records a well observed fluence in the optical, X-ray, gamma-ray and GeV bands, see Fig.~\ref{prompt}.

{The Fermi-GBM count rate of GRB 130427A with isotropic energy $E_{\rm iso}=(9.2 \pm 1.3)\times 10^{53}$~erg and $z=0.34$ is shown in Fig.~\ref{prompt} (a). Clearly identified are a) the supernova raise (SN-rais) \citep{2019arXiv191012615L} b) the UPE phase following the BH formation c) the emission of the cavity mentioned in the introduction, details in Li, et al. (to be submitted). During the UPE phase the event count rate of n9 and n10 of Fermi-GBM surpasses $\sim 8\times 10^4$ counts per second in the prompt radiation between rest-frame times $T_0 + 3.4$~s and $T_0 + 8.6$~s. The GRB is there affected by pile-up, which significantly deforms the spectrum; {details in \citet{2014Sci...343...42A, 2015ApJ...798...10R}}.}

{We therefore impose as the starting point of our analysis the value {$t_{\rm rf}=16$~s,} with $t_{\rm rf}$ being the rest-frame time, and cover all the successive Fermi-GBM and Fermi-LAT data; see Fig.~\ref{fig:sdow}a.}

{In Fig.~\ref{prompt} (b) we give the luminosity of Fermi-LAT (red) and Fermi-GBM (blue), details in \citet{ 2015ApJ...798...10R} and \citet{2019ApJ...878...52A}. From the observations in Fig.~\ref{prompt} (b) at the onset of the GeV emission  the magnetic field $B_0$ and the corresponding electric field are largely overcritical, $E>E_c$ \citep{ 2019arXiv190404162R}. In these conditions, a plasma consisting of a vast number of $e^+e^-$ pairs is produced by the vacuum polarization process. Such a plasma self-accelerates and emits at transparency region the MeV radiation, see e.g. a vast literature quoted in \citet{2007AIPC..910...55R}.}
{The vacuum polarization process create the optically tick condition by which the GeV radiation is drastically reduced until the end of the  UPE phase is reached, see \citep{ 2019arXiv190404162R}.}

{It was already shown in \citet{1975PhRvL..35..463D} that the feedback of such vacuum polarization process can reduce the original overcritical magnetic field down to $\sim 10^{11}$~G.}

{
One of the new issues opened by the data in  Fig.~\ref{prompt} (b), shown in more detailed in Fig.~\ref{prompt} (c),  is precisely the conversion of the GeV photons into the MeV photons for $t_{\rm rf}< 16$~s. The conversion mechanism likely involves the Breit-Wheeler \citep{1934PhRv...46.1087B} photon-photon pair creation $\gamma+\gamma\rightarrow e^+ + e^-$; for details see \citet{2010PhR...487....1R,2016Ap&SS.361...82R}, since for GeV photons their energy is larger than the threshold energy for pair production. Such process is indeed responsible for absorption of GeV emission in some GRBs \citep[see, e.g.,][]{2011ApJ...729..114A}. This process leads to significant production of optically thick $e^+e^-$ plasma and thermalization of high-energy photons at MeV energy. As the luminosity of photons in the MeV energy range decreases approaching  $t_{\rm rf}=16$~s, its number density decreases and consequently the opacity decreases as well. This implies less absorption of GeV photons: indeed, the flux of GeV photons increases.
}

{Based on our recent work about the hard and soft X-ray flares \citep{2018ApJ...869..151R}, the flare in the MeV band around $t_{\rm rf}=$100~s observed in Fig.~\ref{prompt} (b) clearly occurs in the accreting hypernova ejecta that is well outside the conical GeV emission region. This feature is therefore not associated with the GeV emission mechanism treated in this article and since it occurs outside the cone of the GeV emission, these GeV and MeV radiations are not interacting.}

{
In view of the pile-up effect {of GBM data} indicated in Fig.~\ref{prompt} (a) and the absence of accurate data at $t_{\rm rf}<16$~s we will not approach the study of the overcritical field in GRB 130427A in this article. 
}
{We address instead the observation  after $t_{\rm rf}=16$~s, see Fig.\ref{prompt}, where the condition of transparency of the GeV radiation {is reached}. We determine the self-consistent set of parameters which allow the transparency condition to be implemented and the mass and the spin of the BH will be in this context uniquely determined{; see section \ref{sec:5}}.}


\section{Determination of the mass and spin of the BH}\label{sec:2}

In this section we identify the rotational energy of a Kerr BH as the energy source powering the GeV emission at {$t>t_{\rm rf}=16~$s:} consequently, the mass and spin of the BH {have to} be determined. 

\begin{figure}
\centering
\includegraphics[width=1.09\hsize,clip]{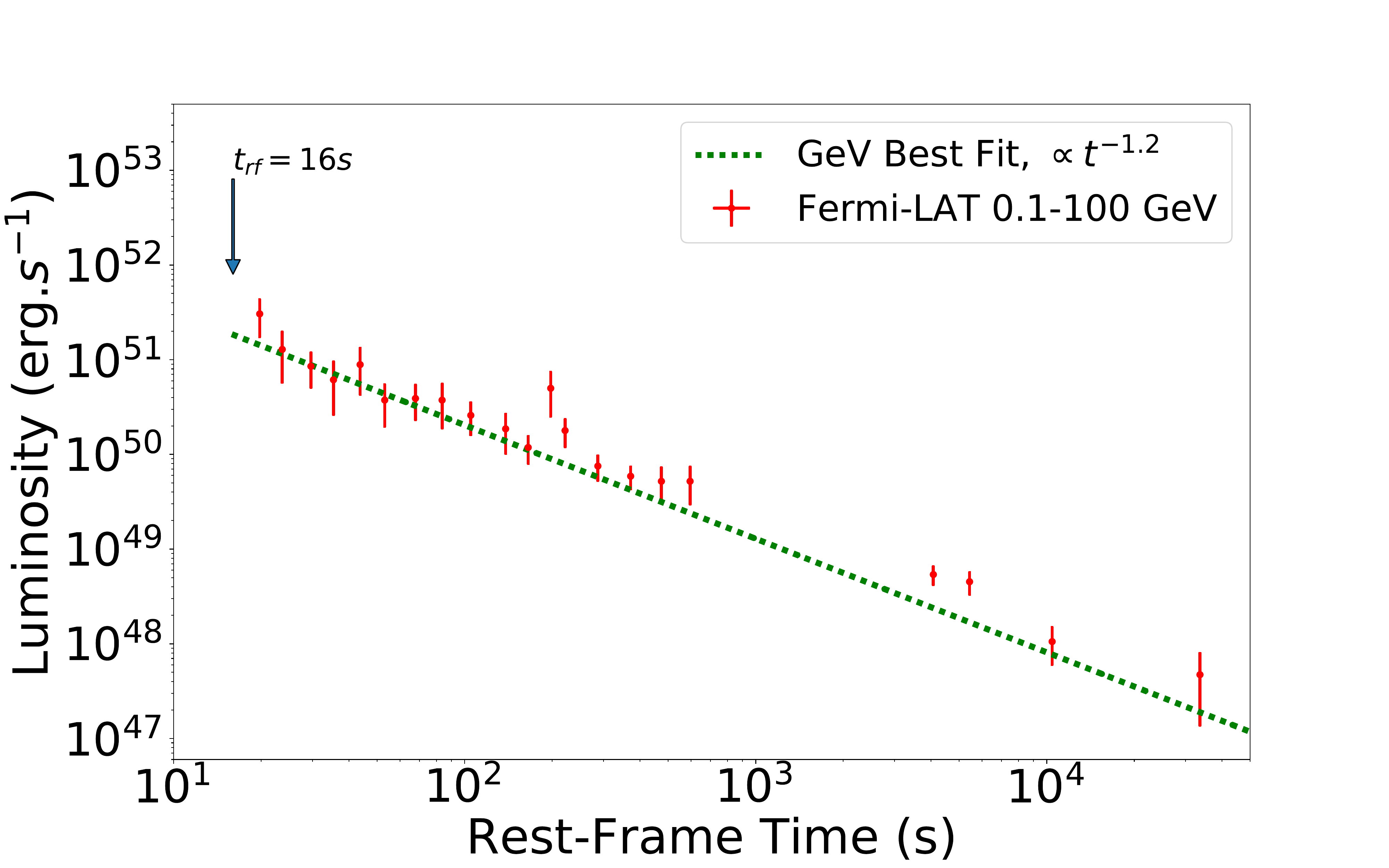}
\caption{ {The rest-frame  $0.1$--$100$~GeV luminosity light-curve of GRB 130427A obtained from \textit{Fermi}-LAT, respectively. The green line shows the best fit for power-law behavior of the  luminosity with slope of $1.2\pm 0.04$ and amplitude of $5.125 \times 10^{52}$~erg~s$^{-1}$.} }
\label{fig:luminosityGeV}
\end{figure}

{The  luminosity of   Fermi-LAT ($0.1$--$100$ GeV) together with a power law best fit to the GeV luminosity of this GRB after $t>t_{\rm rf}=16~$s are shown in Fig.~\ref{fig:luminosityGeV}. }

{After $t>t_{\rm rf}=16~$s,   $E_{\rm GeV}= (1.2 \pm 0.01)\times 10^{53}$~erg, and the GeV luminosity is best fitted by}

\begin{equation}
 L =A~\left(\frac{t}{{1\rm s}}\right)^{-\eta} \rm ~erg~s^{-1},   
 \label{luminosity}
\end{equation}
{with slope of $\eta=1.2\pm 0.04$ and amplitude of $A=(5.125\pm 0.2) \times 10^{52}$, data and energy are retrieved from second Fermi-LAT catalog \citep{2019ApJ...878...52A}.}

We now verify that the energetics of the GeV radiation can be explained by the extractable rotational energy of the Kerr BH, i.e., 
\begin{equation}
\label{Eextr1}
E_{\rm GeV} = E_{\rm extr}=(1.2 \pm 0.01)\times 10^{53}~{\rm erg}.
\end{equation}
{From the mass-energy formula of the Kerr BH (\citealp{1970PhRvL..25.1596C,1971PhRvD...4.3552C,1971PhRvL..26.1344H}; see also ch.33 in \citealp{1973grav.book.....M}) we have}
\begin{subequations}
\renewcommand{\theequation}{\theparentequation.\arabic{equation}}
\begin{align}
\label{aone}
M^2 = \frac{c^2 J^2}{4 G^2 M^2_{\rm irr}}+M_{\rm irr}^2,\\
S = 16\,\pi\,G^2\,M^2_{\rm irr}/c^4,
\end{align}
\end{subequations}
{where  $J$, $M$, $M_{irr}$ and $S$ are the angular momentum, mass, irreducible mass and horizon surface area of the Kerr BH, from which we obtain consequently the extractable energy:} 
\begin{equation}
\label{Eextr}
E_{\rm extr}=M c^2-M_{\rm irr} c^2=\left(1-\sqrt{\frac{1+\sqrt{1-\alpha^2}}{2}}\right)M c^2,
\end{equation}
{where $\alpha = c\,a/(G M) = cJ/(G M^2)$ is the dimensionless angular momentum parameter, being $a=J/M$ the angular momentum per unit mass.}

{
Since we have two unknowns, $M$ and $\alpha$, and only one equation for one observable, Eq.~(\ref{Eextr1}), we need to provide a closure equation to the system to determining the two BH parameters.
} 

{In section~\ref{sec:4}, we show how the transparency condition and  the demand of the synchrotron radiation timescale to be equal to the  timescale of the first impulsive event, inferred from the theory of ``\textit{inner engine}'', gives the additional constraint to determine the mass and spin of the Kerr BH in this GRB.}

\section{On the electrodynamics of the ``inner engine''}\label{innerengine}

We turn now to the electrodynamical mechanism  which extracts the rotational energy in the \textit{inner engine}.

\begin{figure*}
\centering
[a]\includegraphics[width=0.4\hsize,clip]{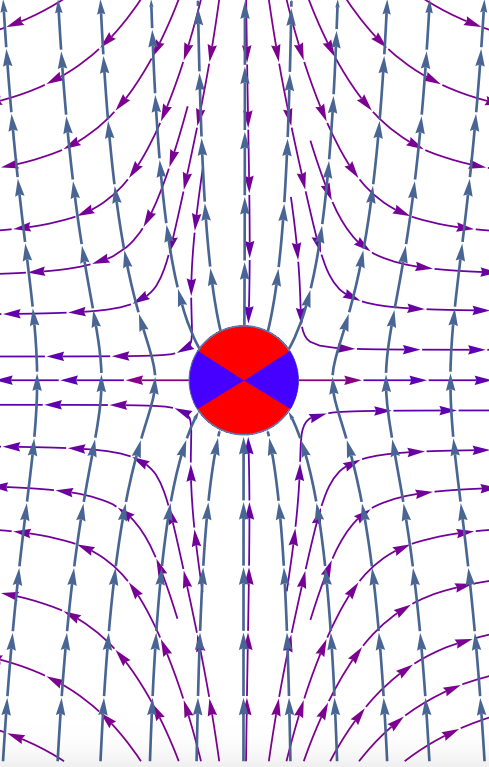}
\vline \vline \vline \vline \vline \vline \vline
[b]\includegraphics[width=0.4\hsize,clip]{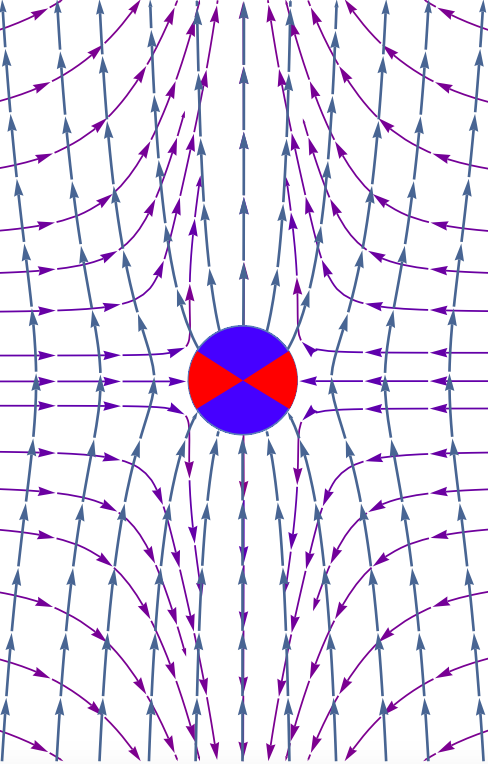}

\caption{The electromagnetic field lines of the Wald solution. The blue lines show the magnetic field lines and the violet show the electric field lines. \textbf{a:} Magnetic field is ``parallel'' to the spin of the Kerr BH, so parallel to the rotation axis. On the polar axis up to $\theta \sim \pi/3$ electric field lines are inwardly directed, therefore electrons are accelerated away from the BH. For $\theta > pi/3 $ electric field lines are outwardly directed and consequently protons are accelerated away from the BH.  \textbf{b:} 
Magnetic field is ``antiparallel'' to the Kerr BH rotation axis.  On the polar axis up to $\theta \sim \pi/3$ electric field lines are outwardly directed, therefore protons will be accelerated away from the BH. For $\theta > \pi/3 $ electric field lines are inwardly directed and consequently  electrons will be accelerated away from the BH.}
\label{energyandpotential}
\end{figure*}

We focus on a Wald solution within a cone of opening angle $\pi/3$ about the magnetic field direction, see Fig.~\ref{energyandpotential}. {We consider in Fig.~\ref{energyandpotential}  the case of magnetic field “parallel”,Fig.~\ref{energyandpotential} (a) (antiparallel, Fig.~\ref{energyandpotential} (b)) to the Kerr BH rotation axis, in which the electrons (protons) are accelerated away in the polar direction.} 

{In this article we shall address the case of magnetic field ``parallel'' to the Kerr BH rotation axis, in which the electrons  are accelerated away in the polar direction, see left figure in  Fig.~\ref{energyandpotential} (a). }

{We address only the leading in the {angular} and radial dependence of the field in the equation of motion. The electromagnetic field of the \textit{inner engine}, in the first-order, slow rotation approximation and at second-order, small angle approximation, reads:
\begin{align}
   E_{\hat{r}} & \approx \frac{a B_0}{r} \left[ \left(1+\frac{G M}{c^2 r} \right)\theta^2 - \frac{2 G M}{c^2 r} \right],\\
    E_{\hat{\theta}}& \approx \frac{a B_0}{r} \left(1 - \frac{2 G M}{c^2 r} \right)^{1/2} \theta,\\
  B_{\hat{r}}& \approx B_0 \left(1-\frac{\theta^2}{2}  \right),\\
  B_{\hat{\theta}}& \approx -B_0 \left(1 - \frac{2 G M}{c^2 r} \right)^{1/2} \theta.
\end{align}
}

{
Up to linear order in $\theta$, the radial component of the electric field can be approximated by the expression
\begin{equation}\label{eq:ER2}
  E_r \approx -\frac{1}{2}\alpha B_0\,c\frac{r_+^2}{r^2}.
\end{equation}
}

{
At the BH horizon, $r_+ = (1+\sqrt{1-\alpha^2}) G M/c^2$, the above electromagnetic field becomes
\begin{align}
   E_{\hat{r}} & \approx -\frac{1}{2}\alpha B_0\,c \left(1-\frac{3}{2}\theta^2\right),\label{eq:Erhorizon}\\
    E_{\hat{\theta}}& \approx 0,\label{eq:Ethetahorizon}\\
  B_{\hat{r}}& \approx B_0 \left(1-\frac{\theta^2}{2}  \right),\label{eq:Brhorizon}\\
  B_{\hat{\theta}}& \approx 0,\label{eq:Bthetahorizon}
 \end{align}
It can be seen from the full numerical solution keeping all orders in the angular momentum, shown in Fig.~\ref{energyandpotential}, that this approximation is valid up to $ \theta_{\pm}= \pi/3$ and for arbitrary value of $\alpha$  and within such limits our small angle approximation gives accurate qualitative and quantitative results.
}

We show how in the presence of a fully ionized low-density plasma, the GRB \textit{inner engine} accelerates electrons up to ultrarelativistic energies in the above mentioned cavity. 
{We assume that the emission process occurs near the BH and within magnetic field lines  constant in time and uniform in space. The equations of motion for the electrons injected for selected angles $\theta$ are given below and specific examples in the section~\ref{sec:4}.}

When emitted in the polar direction $\theta=0$, \textit{the inner engine} can give rise to UHECRs. For $\theta \neq 0$ we integrate the equations of motion and evaluate the synchrotron emission keeping the leading terms.


\section{Synchrotron emission and the first elementary impulsive event}\label{sec:4}

The relativistic expression for the Lorentz force is
\begin{equation}
\frac{dp^{\mu}}{d\tau}=\frac{e}{c}F^{\mu\nu}u_{\nu},\qquad p^{\mu}=mu^{\mu},\qquad u^{\mu}=\frac{dx^{\mu}}{d\tau},
\end{equation}
where $\tau$ is the proper time, $p^\mu$ is the four-momentum, $u^\mu$ is the four-velocity, $x^\mu$ are the coordinates, $F^{\mu\nu}$ is the electromagnetic
field tensor, $m$ is the particle mass, $e$ is the elementary charge and $c$ is the speed of light. This expression can be rewritten in the laboratory frame using vector notation as%
\begin{equation}
mc\frac{d\left(\gamma\mathbf{v}\right)}{dt}=e\left(\mathbf{E}
+\mathbf{v}\times\mathbf{B}\right).
\end{equation}

Assuming the one-dimensional motion along the radial directions,  the dynamics of the electrons in the electromagnetic field (\ref{eq:Erhorizon})-(\ref{eq:Bthetahorizon}), for $\gamma\gg 1$, is determined by the equation
\citep[see, e.g.,][]{1996ApJ...457..253D}

\begin{equation}
m_e c^{2}\frac{d\gamma}{dt}=e\frac{1}{2}\alpha B_{0}\,c^2-\frac{2}{3}e^4\frac{B_0^2\sin^2\left\langle \theta\right\rangle}{m_e^2 c^3}\gamma^2 c^2,\label{eqm}%
\end{equation}
where $\gamma$ is the electron Lorentz factor, $\left\langle \theta\right\rangle$ is the injection angle between the direction of electron motion and the magnetic field and $m_{e}$ is the electron mass. This equation is here integrated for electrons assumed to be injected near the horizon, for selected value of the injection angle $\left\langle \theta\right\rangle$, with an initial Lorentz factor of $\gamma=1$ at $t=0$.

Equation~(\ref{eqm}) is valid for every injection angle $\theta$. The angle dependence in the electric field in Eq.~(\ref{eqm}) is neglected since the second term of the right-hand side of Eq.~(\ref{eqm}), namely the synchrotron radiation term, is largely dominant for the parameters of interest in this work.

Assuming all parameters are constant, the approximate solution in the limit $\gamma\gg1$ is
\begin{align}
 &\gamma=\nonumber\\
 &{\gamma_{\mathrm{max}}\tanh\left[  \frac{2}{3}\frac{e^{2}}{\hbar
c}\left(\frac{B_{0}\sin\left\langle \theta\right\rangle }{B_{c}}\right)
^{2}\gamma_{\mathrm{max}}\frac
{t}{\hbar/m_{e}c^{2}}\right]  ,\label{eqms}}%
\end{align}
which has the following asymptotic value:
\begin{equation}
{\gamma=}\left\{ 
\begin{array}
[c]{cc}%
{\frac{1}{2}\frac{B_{0}}{B_{c}}\alpha\frac
{t}{\hbar/m_{e}c^{2}},} & {t\ll t_{c},}\\
{\gamma_{\mathrm{max}}}, &{ t\gg t_{c},}
\end{array}
\right.  \label{gas}%
\end{equation}
where
\begin{equation}
{\gamma_{\mathrm{max}}=\frac{1}{2}\left(  \frac{3}{\frac{e^{2}}{\hbar
c}}\alpha\frac{B_{c}}{B_{0}\sin^{2}\left\langle \theta\right\rangle
}\right)^{1/2},\label{gmax}}%
\end{equation}
and the critical time is
\begin{equation}
t_c=\frac{\hbar}{m_e c^{2}}\frac{3}{\sin\left\langle \theta\right\rangle
}\left[\frac{e^{2}}{\hbar c}\left(
\frac{B_0}{B_c}\right)^{3}\alpha\right]^{-1/2}.%
\label{tcr}
\end{equation}

The maximum {peak photon energy of the synchrotron spectrum} is obtained {by using the maximum Lorentz factor of the radiating electrons which is given by the} equilibrium between energy gain and energy loss in Eq.~(\ref{eqm}). Consequently, the following maximum energy of the electron-synchrotron photons is found: 
\begin{align}\label{maxgev}
{\epsilon_{\mathrm{max,\gamma}}}&={\frac{3e\hbar}{2m_{e}c}B_{0}\sin\left\langle
\theta\right\rangle \gamma_{\mathrm{max}}^{2}=\frac{9}{8}\frac
{m_{e}c^{2}}{e^{2}/\hbar c}\frac{\alpha}{\sin\left\langle \theta
\right\rangle}} \nonumber\\ 
&{\approx\frac{80}{\sin\left\langle \theta\right\rangle
}\alpha\,\mathrm{MeV}.}
\end{align}
The maximum energy is independent of the magnetic field strength, which for different angles leads to different energy bands for the photons{; see section~\ref{sec:power}}. From this upper limit some inferences on the TeV emission are in preparation, awaiting the publication of the TeV data. Here we return to the GeV emission and to its energy originating from the BH rotational energy.

A vast literature exists on the propagation of ultra high energy protons/electrons in a magnetic field with $\theta=\pi/2$ \citep[see e.g.,][and references therein]{1966RvMP...38..626E}. Very little has been published for computations for small injection angles, $\theta\approx 0$ (an important exception being \citealp{1991Sci...251.1033H}) which we are also here addressing.

{The maximum electric potential difference associated with the electric field is obtained by bringing an electron from the BH horizon to infinity along the symmetry/rotation ($\theta=0$) axis:
\begin{eqnarray}\label{eq:deltaphi}
 \Delta \phi &=&  \frac{\epsilon_e}{e} = \int_{r_+}^\infty E dr = E_{r_+} r_+ \nonumber \\
    &=& 9.7\times 10^{20}\cdot \alpha \beta \mu (1+\sqrt{1-\alpha^2})\quad\frac{{\rm eV}}{e},
\end{eqnarray}
where we have introduced $\beta\equiv B_0/B_c$ and $\mu\equiv M/M_\odot$, and $E_{r_+}$ is the electric field evaluated at the horizon (see Eq.~\ref{eq:Erhorizon})
\begin{equation}\label{eq:Eh}
    E_{r_+}={-}\frac{1}{2}\alpha B_0 c.
\end{equation}
This potential can accelerate electrons along the symmetry axis up to a maximum Lorentz factor and energy given by:
\begin{equation}
\label{epemax}
\epsilon_{e,max}=e \Delta \phi=\gamma m_e c^2.
\end{equation}
For $\theta=0$ there is no energy loss due to synchrotron radiation, hence  the total electrostatic energy goes into electron acceleration. The time of acceleration for $\theta=0$ can be obtained from Eq.~(\ref{eqm}) when the synchrotron loss term in the right-hand side is zero, i.e.
\begin{equation}\label{eqmm}
m_e c^{2}\frac{d\gamma}{dt}=e\frac{1}{2}\alpha B_{0}\,c^2=e  E_{r_+} c,%
\end{equation}
therefore,
\begin{equation}
t(\theta=0)=\frac{ m_e c^{2} \gamma}{e E_{r_+} c}= \frac{E_{r_+} r_+}{E_{r_+} c}=\frac{r_+}{c},%
\end{equation}
where we have used Eqs.~(\ref{eq:deltaphi}) and (\ref{epemax}).}

{\subsection{Timescale of the first impulsive event}}

The electrostatic energy available is
\begin{equation}
{\cal E} =\frac{1}{2} E_{r_+}^2 r_+^3= 7.5\times 10^{41}\cdot\alpha^2 \beta^2 \mu^3 (1+\sqrt{1-\alpha^2})^3~ \rm erg,
\label{eq:em}
\end{equation}
where we have used Eq.~(\ref{eq:Eh}).

Therefore, the timescale of the first impulsive event {obtained from the  GeV luminosity at $t_{\rm rf}=16$~s, denoted as ${\cal E}_1\equiv {\cal E_{\rm t_{\rm rf}=16{\rm s}}}$, is}
\begin{equation}
{{ \tau_{\rm 1} =\frac{{\cal E_{\rm t_{\rm rf}=16{\rm s}}}}{L_{\rm GeV}({t_{\rm rf}=16s})}}}
\label{timescalew}
\end{equation}
{which reads:
\begin{equation}
\tau_{\rm 1} =4.08\times 10^{-10}  \alpha^2 \beta^2  \mu^3 (1+\sqrt{1-\alpha^2})^3\,{\rm s},
\label{timescalew2}
\end{equation}
where we have used Eq.~(\ref{luminosity}).}

{We turn now to a crucial relation between $\alpha$ and $\beta$.}

{\subsection{Transparency of GeV photons}\label{sec:5}}

{The hypercritical accretion onto the NS and its subsequent collapse forming the BH, deplete the BdHN by $\approx 10^{57}$ baryons, creating a cavity of $\approx 10^{11}$~cm of radius in the hypernova ejecta around the BH site \citep[see][]{2016ApJ...833..107B,2019ApJ...871...14B}. The density inside the cavity at BH formation is about $10^{-6}$~g~cm$^{-3}$ \citep{2016ApJ...833..107B,2019ApJ...871...14B} and it is further decreased to about $10^{-13}$~g~cm$^{-3}$ by the GRB explosion \citep{2019ApJ...883..191R}. The low density of this cavity guarantees a condition of low baryon density necessary for the transparency and so for the observation of the MeV emission in the UPE, as well as the higher energy band emission discussed in the present work \citep[see][for details]{2019ApJ...883..191R,2019arXiv190404162R}.}

\begin{figure}[ptbh]
\centering
\includegraphics[width=.45\textwidth]{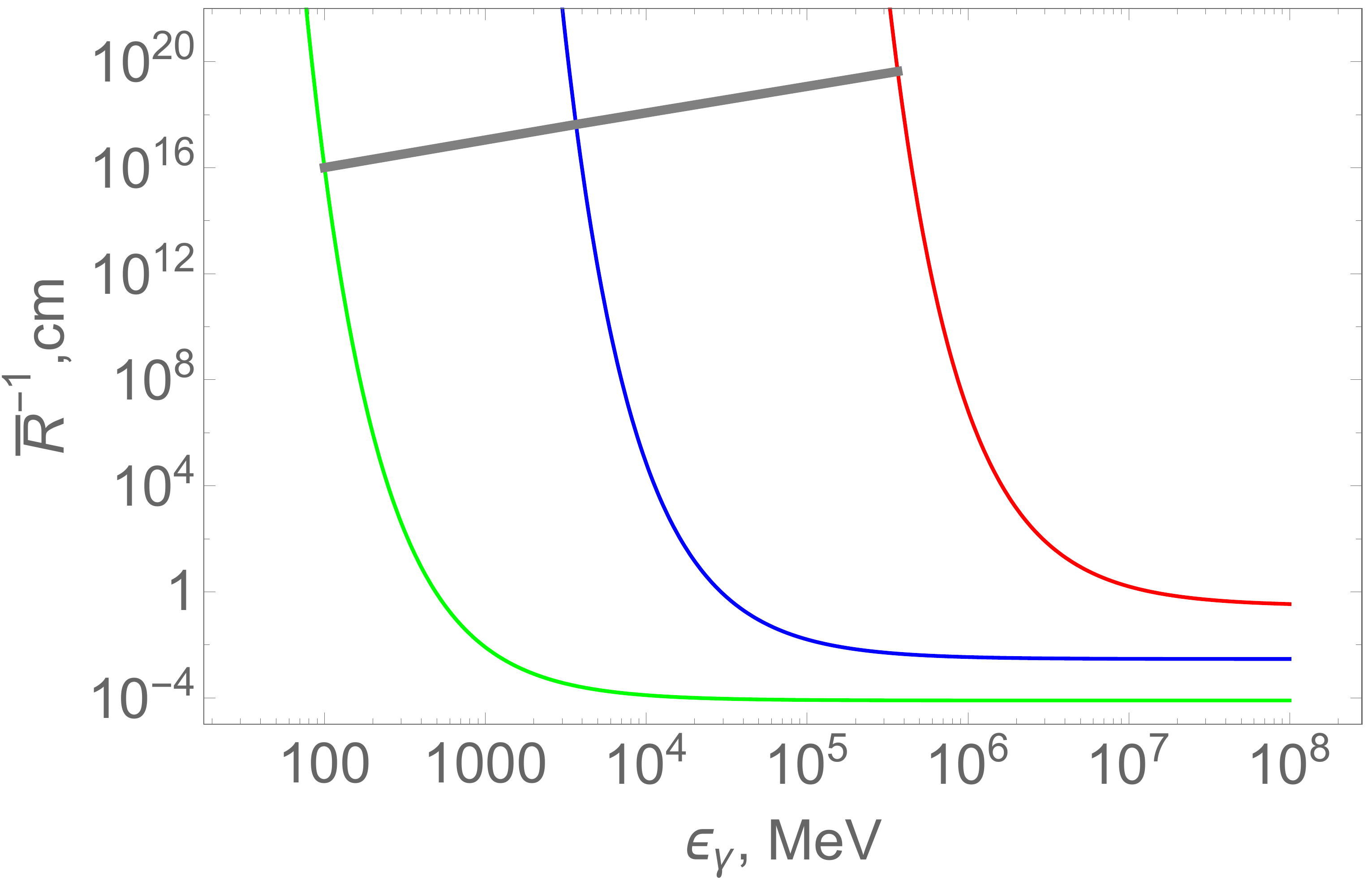} \caption{{Inverse of the attenuation
coefficient for pair production in magnetic field, Eq.~ (\ref{Batteq}) computed for $\beta=B_0/B_c=7.84\times10^{-4}$ and selected $\left\langle \theta\right\rangle$ of $\pi/8$ (green),~ $10^{-2}$ (blue) and ~$10^{-4}$ (red). The peak of the synchrotron spectrum, given by Eq.~(\ref{maxgev}) and shown by the gray line, is in the transparent region.} }%
\label{Batt}%
\end{figure}

{But this condition for transparency, necessary, is not sufficient. There is a most stringent condition imposed by the interaction of the synchrotron photons with the field $B_0$. Synchrotron photons of energy $\epsilon_\gamma$ may produce $e^+e^-$ pairs in external magnetic field. The inverse of the attenuation coefficient \citep{1983ApJ...273..761D}}
\begin{equation}
{\bar{R}\sim0.23\frac{e^{2}}{\hbar c}\left(  \frac{\hbar}{m_{e}c^{2}}\right)
^{-1}\beta\sin\left\langle \theta\right\rangle\exp\left(
-\frac{4/3}{\frac{\epsilon_\gamma}{2m_ec^2}\beta\sin\left\langle
\theta\right\rangle}\right).}\label{Batteq}%
\end{equation}
{Imposing the following transparency condition for $0.1$~GeV photons, $\bar{R}^{-1} \geq 10^{16}$~cm, we obtain:}

\begin{equation}\label{eq:Btransparency}
  {  \beta \leq 3.67 \times 10^{-4} \alpha^{-1},}
\end{equation}
{
where we have replaced Eq.~(\ref{maxgev}) in (\ref{Batteq}) to express the dependence of $\bar{R}$ on the pitch angle in terms of the peak photon energy and the spin parameter; {see Fig.~\ref{Batt}}.
}

{We shall use this equation in next section in order to obtain mass and spin of BH in GRB 130427A.}

{\section{Mass and spin of BH}\label{sec:massspin}}

{
Since there are three unknown parameters, namely $M$, $\alpha$ and $\beta$, and only two equations, namely Eq.~(\ref{Eextr1}) via Eq.~(\ref{Eextr}), and Eq.~(\ref{eq:Btransparency}), an additional equation is needed to determine the three parameters of the \textit{inner engine}.
}

{For this purpose we request the additional constraint that the timescale of the synchrotron radiation, obtained from Eq.~(\ref{tcr}), be equal to the radiation timescale obtained from Eq.~(\ref{timescalew}), at the time $t_{\rm rf}=16~$~s, i.e. the following three equations must be solved simultaneously:
\begin{eqnarray}
 E_{\rm GeV} &=& E_{\rm extr}(\mu,\alpha)\label{eq:sys1}\\
  \beta &=& 3.67 \times 10^{-4} \alpha^{-1}\\
 t_c (\left\langle\theta\right\rangle,\alpha,\beta)&=& \tau_{\rm 1}(\mu,\alpha,\beta,L_{\rm GeV}).\label{eq:sys2}
\end{eqnarray}
We can solve this system of equations as follows. First, from Eq.~(\ref{eq:sys1}) we can isolate $\mu$ as a function of $E_{\rm GeV}$ and $\alpha$:
\begin{equation}\label{eq:mu}
    \mu = \left(1 - \sqrt{\frac{1 + \sqrt{1 - \alpha^2}}{2}}\right)^{-1}\,\frac{E_{\rm GeV}}{M_\odot c^2}.
\end{equation}
Then, by replacing Eq.~(\ref{eq:mu}) into Eq.~(\ref{eq:sys2}), we obtain an expression for $\beta$ as a function of the observables $E_{\rm GeV}$ and $L_{\rm GeV}$, and $\alpha$:
\begin{eqnarray}\label{eq:alphabeta}
    \beta&=&\beta(\epsilon_\gamma,E_{\rm GeV},L_{\rm GeV},\alpha) \nonumber \\
    &=& \frac{1}{\alpha}\left(\frac{64}{9} \sqrt{3\frac{e^2}{\hbar c}}\frac{\epsilon_\gamma}{B_c^2\,r_+(\mu,\alpha)^3}\frac{L_{\rm GeV}}{e\,B_c\,c^2}\right)^{2/7},
\end{eqnarray}
where we have replaced Eq.~(\ref{maxgev}) into Eq.~(\ref{tcr}) to express $t_c$ as a function of the peak photon energy $\epsilon_\gamma$instead of the pitch angle, and $r_+$ is the BH horizon which is a function of $\mu$ and $\alpha$ but, via Eq.~(\ref{eq:mu}), becomes a function of $E_{\rm GeV}$ and $\alpha$.}

\begin{figure}
    \centering
    \includegraphics[width=\hsize,clip]{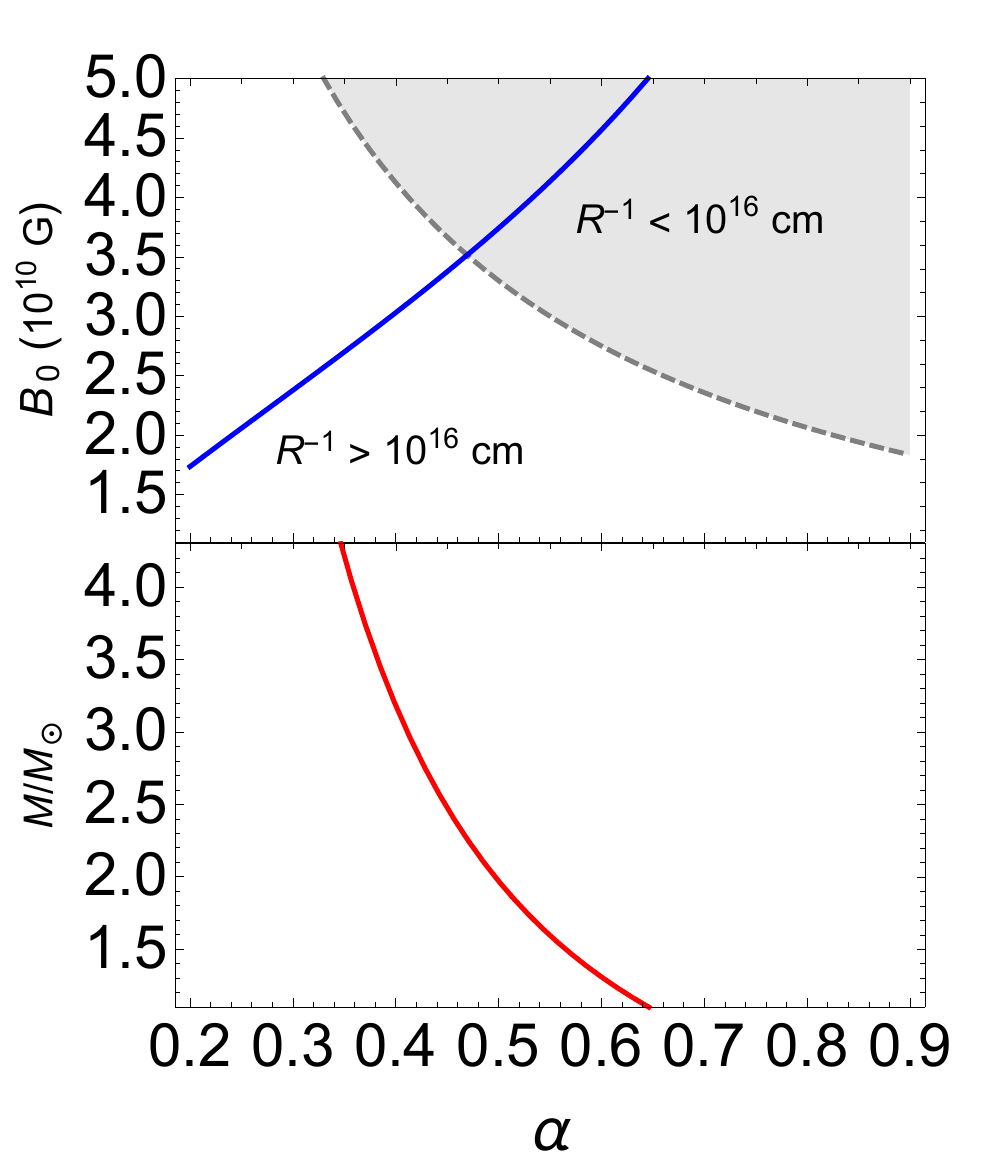}
    \caption{{Upper panel: self-consistent family of solutions of the magnetic field $B_0$ as a function of the BH spin parameter $\alpha$ (blue curve), Eq.~(\ref{eq:alphabeta}), for given values of $E_{\rm GeV}=1.2\times 10^{53}$~erg, $L_{\rm GeV}=1.84\times 10^{51}$~erg~s$^{-1}$ and $\epsilon_\gamma = 0.1$~GeV. Inside the gray shaded region $0.1$~GeV photons have $\bar{R}^{-1} < 10^{16}$~cm, while in the white one  they fulfill the condition of transparency $\bar{R}^{-1} \geq 10^{16}$~cm, namely Eq.~(\ref{eq:Btransparency}). The crossing between the blue curve and the border of the gray region gives us the upper limit of the magnetic field $B_0\approx 3.5\times 10^{10}$~G and the spin parameter $\alpha=0.47$, to have transparency of the GeV photons. Lower panel: self-consistent solution of the BH mass as a function of the BH spin parameter $\alpha$ (red curve). To the maximum spin parameter for transparency it corresponds a lower limit to the BH mass, $M=2.3 M_\odot$.}}
    \label{fig:Bvsalpha}
\end{figure}

{
Therefore, given the  energy $E_{\rm GeV}$ (integrated for times $t_{\rm rf}\geq 16~$~s) and luminosity $L_{\rm GeV}$ (at $t_{\rm rf}=16~$~s), Eq.~(\ref{eq:alphabeta}) gives the self-consistent family of solutions of the magnetic field $\beta$ as a function of the BH spin parameter $\alpha$. In Fig.~\ref{fig:Bvsalpha} (blue curve in the upper panel) we show such a family of self-consistent solutions in the case of $E_{\rm GeV}=1.2\times 10^{53}$~erg (see Eq.~\ref{Eextr1}), $L_{\rm GeV}=1.84\times 10^{51}$~erg~s$^{-1}$ given by Eq.~(\ref{luminosity}) and photon energy $\epsilon_\gamma = 0.1$~GeV.}

{
From Eq.~(\ref{eq:Btransparency}) we know $\beta$ as a function of $\alpha$ for which the condition of transparency is satisfied. Therefore, by equating Eqs.~(\ref{eq:alphabeta}) and (\ref{eq:Btransparency}) we obtain, as can be seen from Fig.~\ref{fig:Bvsalpha}, a maximum spin parameter $\alpha$ to fulfill the transparency condition for $0.1$~GeV photons. Correspondingly, there is the maximum magnetic field value which can be obtained by substituting the upper value of $\alpha$ either into Eq.~(\ref{eq:alphabeta}) or (\ref{eq:Btransparency}). Then, the maximum $\alpha$ is used in Eq.~(\ref{eq:mu}) to obtain the corresponding lower limit for the BH mass. For the above numbers, the upper magnetic field value to have transparency is $\beta=7.8\times 10^{-4}$, i.e. $B_0\approx 3.5\times 10^{10}$~G. The maximum spin and a minimum BH mass are, respectively, $\alpha\approx 0.47$ and $M=2.3~M_\odot$. For the above spin value, we obtain from Eq.~(\ref{maxgev}) the pitch angle to emit $0.1$~GeV photons,  $\theta \approx \pi/8$. The corresponding BH irreducible mass is $M_{\rm irr}=2.2~M_\odot$ which is close to critical mass of the NS for some specific nuclear equation of state, in particular to the TM1 one \citep[see][]{Cipolletta:2016yqv}. These parameters will be used in next subsection as the initial values of mass and spin parameters to find the spin-down of the BH.
}

{The inverse of the attenuation coefficient from Eq.~(\ref{Batteq}) computed for $\beta=7.84\times10^{-4}$ and different $\left\langle \theta\right\rangle$ is presented in Fig.~\ref{Batt}. The peak of the synchrotron spectrum, Eq.~(\ref{maxgev}), shown by the gray line in Fig.~\ref{Batt}, is located in the transparent region. We can see from the figure that synchrotron photons, when produced in the $0.1$~GeV to $1$~TeV energy band, do not produce pairs if magnetic field is below $B_{0}<3.5\times10^{10}$~G. Therefore, this region is transparent for such photons.}

{\subsection{The decrease of the mass and spin of the BH as a function of the extracted rotational energy}\label{sec:spindown}}

From the luminosity expressed in the rest-frame of the sources, and from the initial values of the spin and of the mass of the BH we can  now  derive the slowing down of the BH due to the energy loss in the GeV emission. The time derivative of Eq.~(\ref{Eextr}) gives the luminosity
\begin{equation}
\label{sdown1}
{L=-\frac{dE_{extr}}{dt}=-\frac{dM}{dt},}
\end{equation}
Since $M_{irr}$ is constant for each BH during the energy emission process, and using our relation for luminosity  {from Eq.~(\ref{luminosity})}, we obtain the relation of the loss of mass-energy of the BH by integrating Eq.~(\ref{sdown1}):
\begin{equation}
\label{sdown2}
M= M_0 + 5 A t^{-0.2}- 5A t_0^{-0.2},
\end{equation}
{where $M_0$ is the initial BH mass at, $t_0=16$~s and $A=(5.125\pm 0.2) \times 10^{52}$.} From the mass-energy formula of the BH we have
\begin{equation}
\label{sdown3}
J= 2 M_{irr} \sqrt{M^2-M^2_{irr}},
\end{equation}
therefore
\begin{equation}
\label{sdown4}
a=\frac{J}{M}= 2 M_{irr} \sqrt{1-\frac{M^2_{irr}}{(M_0 + 5 A t^{-0.2}-5A t_0^{-0.2})^2}}.
\end{equation}

\begin{figure*}
\centering
\gridline{\fig{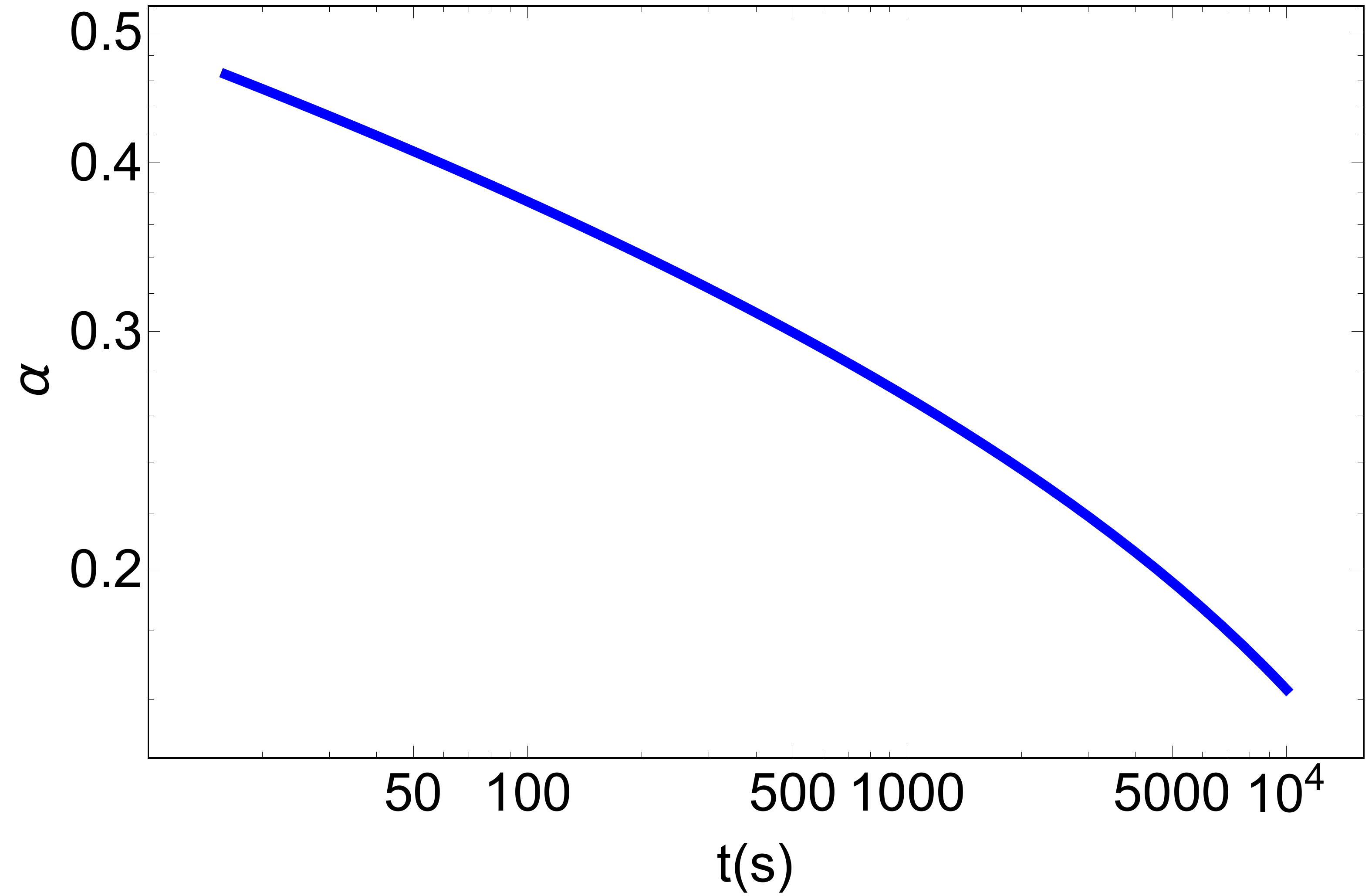}{0.48\textwidth}{(a)}
\fig{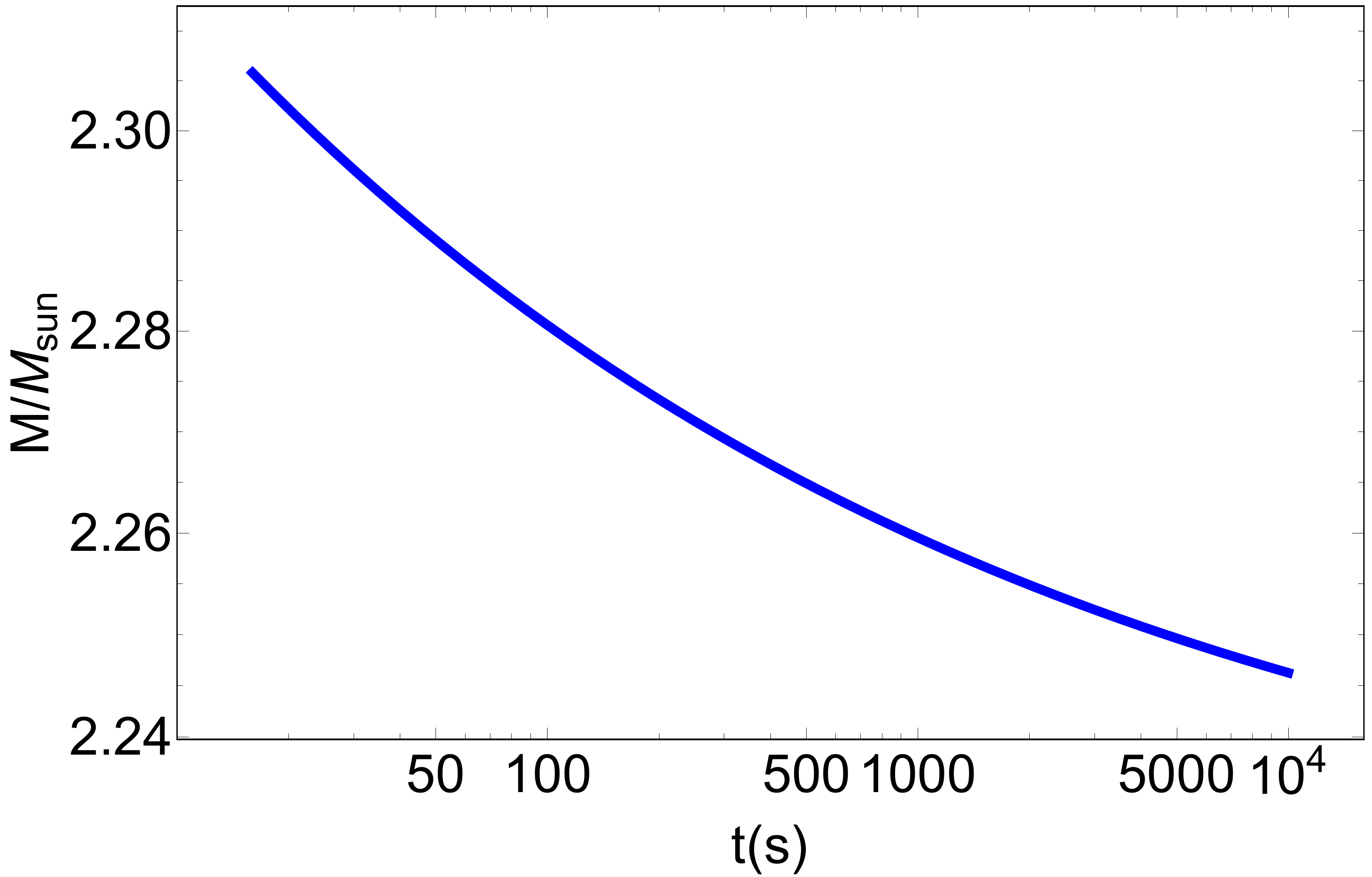}{0.48\textwidth}{(b)}}
\caption{ \textbf{a and b:} The decrease of the BH spin and Mass, as a function of rest-frame time for GRB 130427A. The values of spin and mass at the moment which prompt is finished which has been assumed to occur at the  rest-frame time of $t_{\rm rf}=16$~s, are: $\alpha=0.47$ and $M(\alpha)=2.3 M_{\odot}$.}
\label{fig:sdow}
\end{figure*}

{The values of mass and spin parameters at $t_0=t_{\rm rf}=16$~s; see Fig.~\ref{fig:luminosityGeV} are $M_0=2.3 M_\odot$ and $\alpha = 0.47$ and the irreducible mass is $M_{\rm irr}=2.24~M_\odot$.} The behaviour of $\alpha=J/M^2$ and $M$ with time are shown in Fig.~\ref{fig:sdow}. Both $\alpha$ and $M$ decrease with time which shows the decrease of rotational energy of the BH due to the energy loss in GeV radiation; see Fig.~\ref{fig:sdow}. {It is important to recall that, since we are here inferring the BH energy budget using only the GeV emission data after the UPE phase, the above BH mass and spin have to be considered as lower limits.}

{\section{Synchrotron radiation power and the need of a low density ionizes plasma}} \label{sec:power}

{Having obtained the values of spin, $\alpha=0.47$, mass $M=2.3 M_\odot$ and magnetic field, $B_0/B_c=7.8\times 10^{-4}$, we integrate the equation of motion given by Eq.~(\ref{eqm}) to obtain the radiation in the GeV and TeV  bands corresponding to selected values of $\theta$.}

\begin{figure*}[ptbh]
\centering 
\gridline{\fig{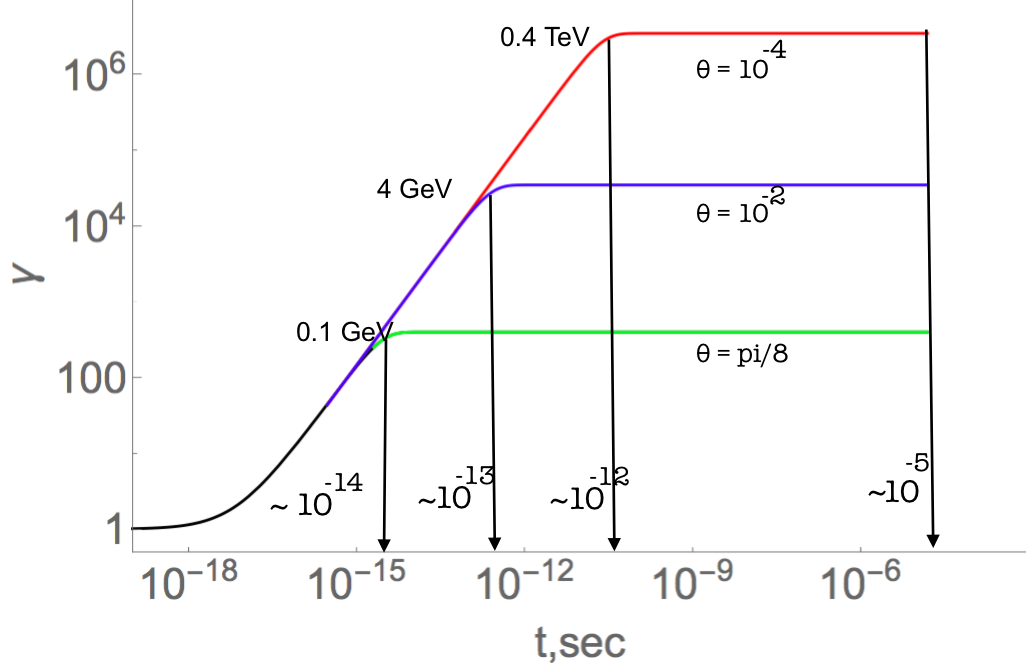}{0.48\textwidth}{(a)}
\fig{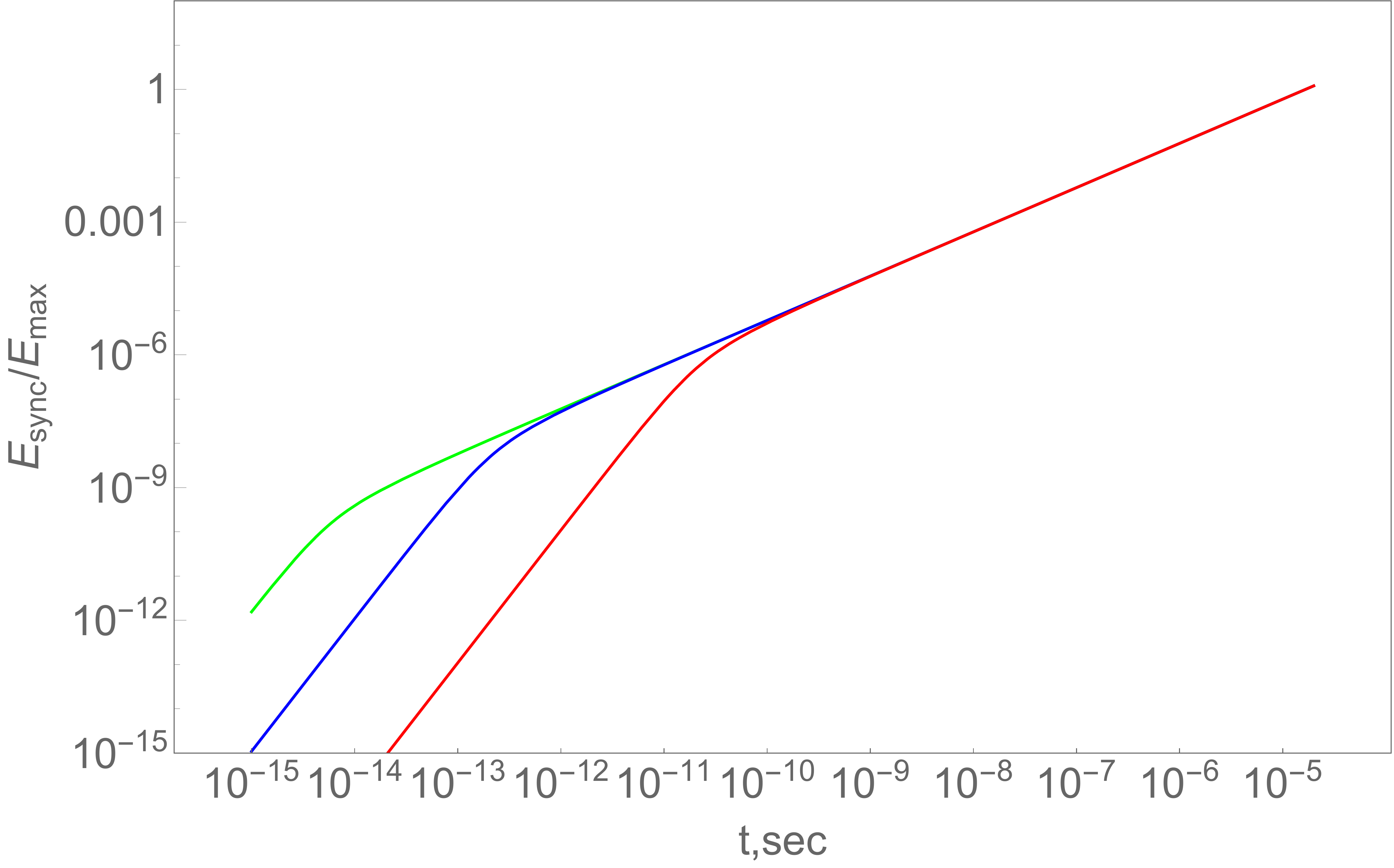}{0.48\textwidth}{(b)}}
\gridline{\fig{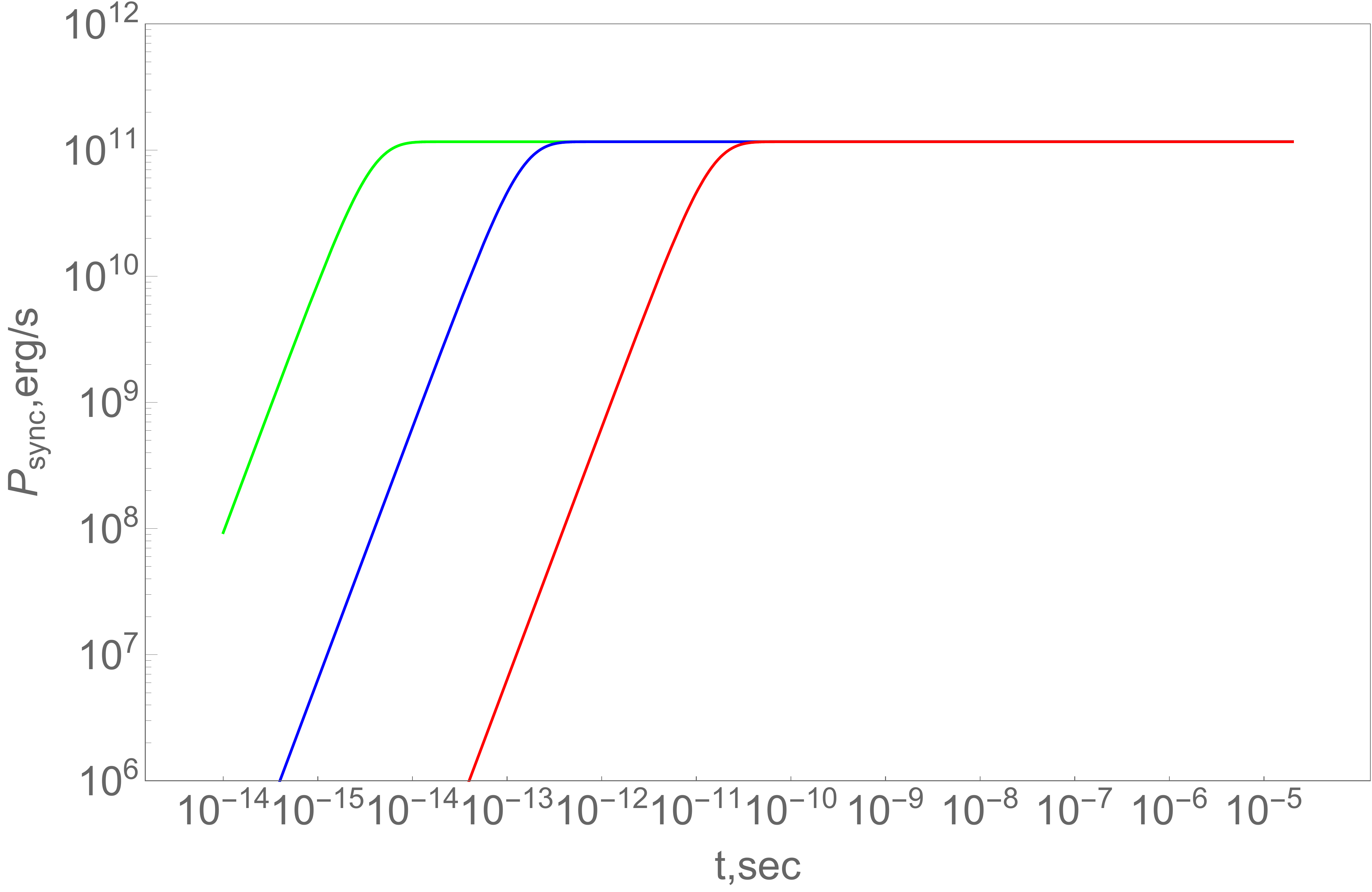}{0.48\textwidth}{(c)}
\fig{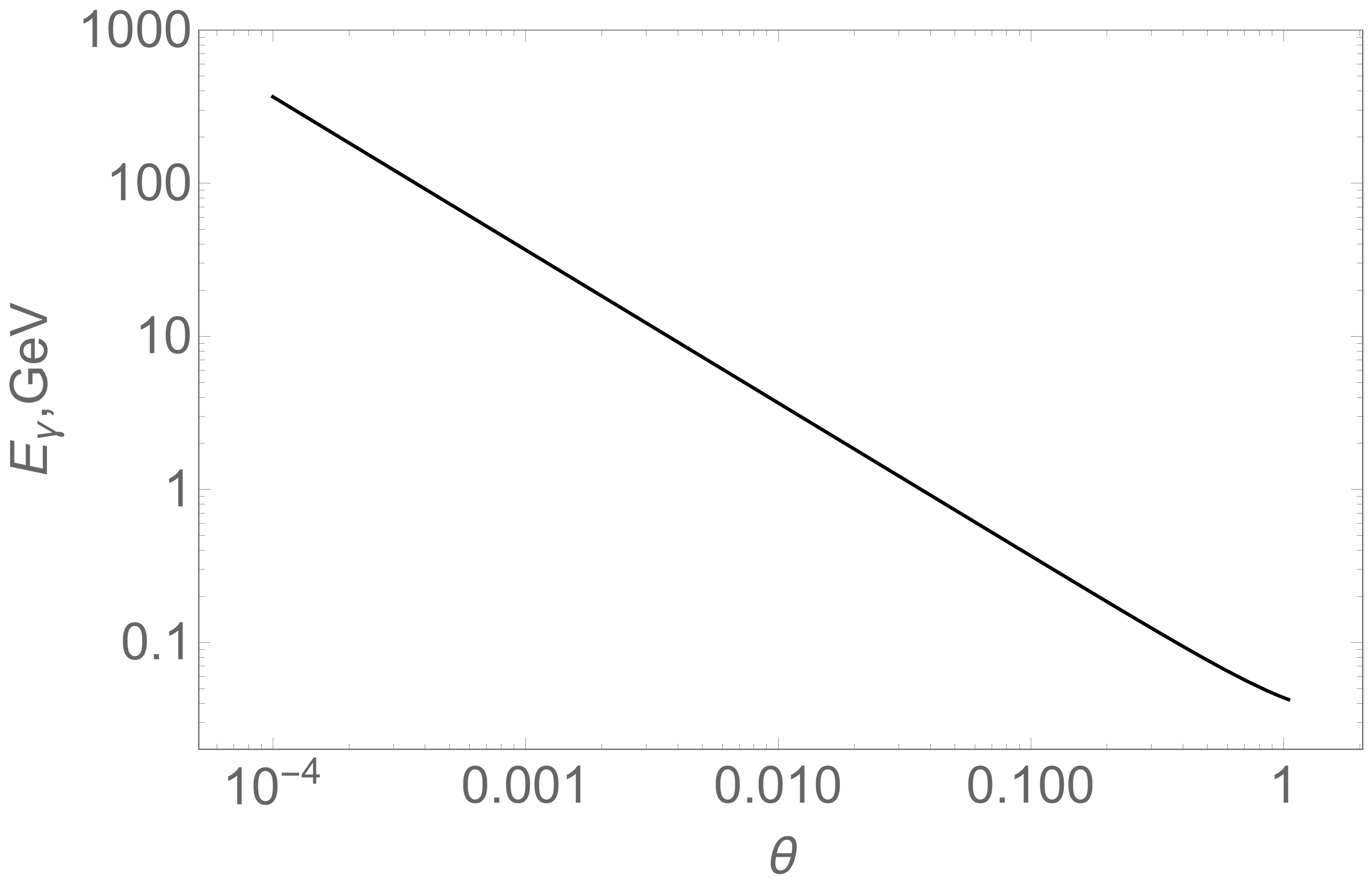}{0.48\textwidth}{(d)}}
\caption{\textbf{a:} The electron Lorentz gamma factor obtained from solutions of Eq.~(\ref{eqm}) as functions of time: numerical (black), and analytic for selected angles: $\theta=\pi/8$ (green), $\theta=1\times10^{-2}$ (blue),
$\theta=1\times10^{-4}$ (red) {which according to Eq.~(\ref{maxgev}),  they are related to $\sim$ 0.1 GeV, $\sim$ 4 GeV and $\sim$ 0.4 TeV energy of synchrotron photons, respectively}. Parameters assumed: $a/M=0.47$, $B_0/B_{c}=7.8 \times 10^{-4}$. The arrow indicates the time when the energy emitted in synchrotron radiation equals $\epsilon_{max,\gamma}=10^{18}$~eV, which is $ t=r_+/c \sim 2 \times 10^{-5}$~s.
\textbf{b:} Total energy emitted in synchrotron radiation from Eq.~(\ref{Esyncsol}) as a function of time for selected angles given in Fig.~\ref{referee1}(a).
\textbf{c:} Power of synchrotron emission by a single {electron} as a function of time for selected angles given in Fig.~\ref{referee1}(a). 
\textbf{d:} Peak energy of synchrotron photons as a function of the angle between the {electron} velocity and the magnetic field.}%
\label{referee1}%
\end{figure*}

\begin{figure*}[ptbh]
\centering
\gridline{\fig{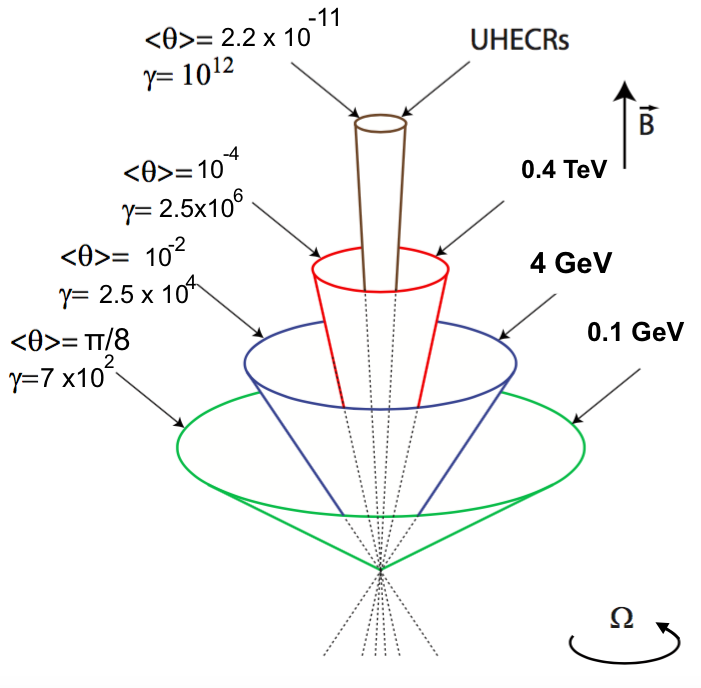}{0.66\textwidth}{(a)}}
\gridline{\fig{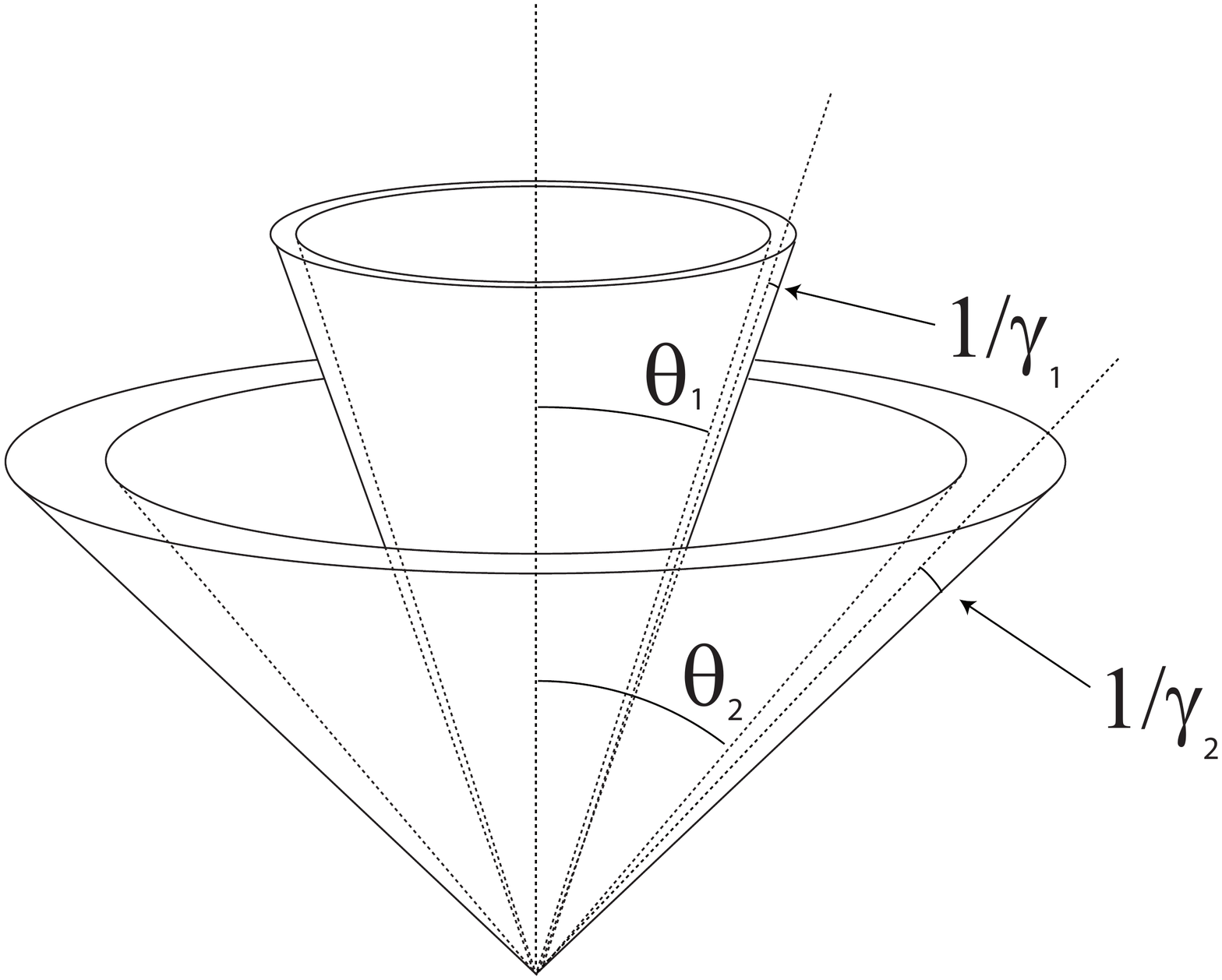}{0.5\textwidth}{(b)}}
\caption{(Not to scale) 
\textbf{a:} Having the values of spin and magnetic field, {$a/M=0.47$, $B_{0}/B_{c}=7.8 \times 10^{-4}$,} from Eq.~(\ref{maxgev}) for selected injection  angles we obtain the radiation in the different bands, {0.1 GeV to 0.4 TeV}. {Using Eq.~(\ref{maxgev}) the maximum energy $\epsilon_{e}\sim 1.6 \times 10^{18}$~eV is reached at the critical angle $\left\langle \theta\right\rangle =2.2\times10^{-11}$.}
This angle is an absolute lower limit for emitting synchrotron radiation, therefore for $\left\langle\theta\right\rangle <2.9\times10^{-11}$, electrons are accelerated to give rise to UHECRs. In this figure the magnetic field is ``parallel'' to the Kerr BH rotation axis and electrons are accelerated outward and protons captured by the horizon.
\textbf{b:} Synchrotron emission from electrons with pitch angle $\theta$. Radiation is concentrated in a cone of angle of $1/\gamma$.}
\label{cones}%
\end{figure*}

As an example we show the results for the electron propagation and radiation for selected angles, i.e., {$\theta=\pi/8$, $\theta=1\times10^{-2}$, $\theta=1\times10^{-4}$} with respect to the direction of the magnetic field. {According to Eq.~(\ref{maxgev}), these angles are related to $\sim$ 0.1 GeV, $\sim$ 4 GeV and $\sim$ 0.4 TeV energy of photons, respectively, which covers the lower limit of Fermi-LAT instrument till the lower limit of MAGIC telescope, see Fig.~\ref{referee1} (a).} The numerical solution of Eq.~(\ref{eqm}) along with analytic solutions are represented in Fig.~\ref{referee1}a. The electron synchrotron luminosity from the right-hand side of Eq.~(\ref{eqm}) is
\begin{equation}
\dot{E}_{\rm sync}=\frac{2}{3}%
e^{4}\frac{B_{0}^{2}\sin^{2}\left\langle \theta\right\rangle }{m_{e}^{2}c^{3}%
}\gamma^{2}.\label{eqm1}%
\end{equation}

In Fig.~\ref{referee1}c we present the total power of the synchrotron emission by a single electron as a function
of time, for selected injected angles $\theta$. The power increases with time and then at $ t>t_c$ approaches a constant value, which does not depend on the angle
\begin{equation}
\dot{E}_{{\rm sync},\gamma_{\rm max}}=\frac{1}{2}\frac{(m_{e}c^{2})^2}{\hbar}\frac{B_{0}}{B_{c}}\alpha={1.16\times 10^{11}\;\text{erg~s}^{-1}.}
\end{equation}

In Fig.~\ref{referee1}d we show the peak energy of the synchrotron photons, given by Eq.~(\ref{maxgev}), as a function of the electron injection angle and the magnetic field.

The total energy emitted by an electron in synchrotron radiation, computed by integrating the synchrotron power with time, gives
\begin{align}
&E_{\rm sync}=\nonumber \\
&{\left\{
\begin{array}
[c]{cc}%
\frac{1}{18}\frac{e^{2}}{\hbar c}\left(\frac{B_0}{B_c}\right)
^{4}\alpha^2\left(\frac{t}{\hbar/m_e c^2}\right)^3\sin^2\left\langle \theta\right\rangle m_e c^2, & t < t_{c},\\
\frac{1}{2}\alpha\frac{B_0}{B_c}\frac
{t}{\hbar/m_e c^2}m_e c^2, & t > t_c,
\end{array}
\right.}\label{Esyncsol}%
\end{align}
where from Eq.~(\ref{tcr}) the critical time for {$\alpha=0.470$, $B_{0}/B_{c}=7.8 \times 10^{-4}$ is}
\begin{equation}\label{tcsynch}
   t_{c} \simeq {\frac{5.9\times10^{-15}}{\sin\left\langle \theta\right\rangle }\,\text{s.}}
\end{equation}

{The synchrotron time scale, obtained from Eq.~(\ref{tcsynch}), has values between $1.6\times 10^{-14} $~s to $2 \times 10^{-5}$~s, see Fig.~\ref{referee1} (a). According to Eq.~(\ref{Esyncsol}) the total energy available for the synchrotron radiation of one electron is a function of synchrotron time scale, see Fig.~\ref{referee1} (b). Each timescale corresponds to the time which takes for one electron to radiate by synchrotron mechanism its acceleration energy $\epsilon_e=\gamma_{\rm max} m_e c^2$, where $\gamma_{\rm max}$ is given by Eq.~(\ref{gmax}). For $\theta = \pi/8$ which is related to the lower limit of Fermi-LAT energy bandwidth, namely $0.1$~GeV, $t_c= 1.6\times 10^{-14}$~s corresponds to $\epsilon_e = 1.02 \times 10^{9}$~eV. Therefore, }
\begin{equation}\label{eqm2}
{{t_{\rm c,GeV}=1.6\times 10^{-14}\,\text{s}.}}
\end{equation}

{For the available electromagnetic energy budget the system can accelerate a total number of electrons with the above energy:}
\begin{equation}\label{enumber}
{N_e = {\cal E_{\rm 1}}/\epsilon_e \approx 6.6\times 10^{39}.}
\end{equation}
{where ${\cal E}_1\approx 10^{37}$~erg is the electrostatic energy available, the blackholic quantum \citep{2019arXiv190708066R}, for the first impulsive event obtained from Eq.~(\ref{eq:em})}. 

{In principle as the timescale increases the total available electron will decrease, we will return to this subject in section~\ref{sec:6} when we study the evolution of the time of sequence of elementary impulses.}

Using Eq.~(\ref{maxgev})  the maximum energy $ \epsilon_{e}\sim 1.6 \times 10^{18}$~eV is reached at the critical angle ${\left\langle \theta\right\rangle \approx 2.2 \times10^{-11}}$. This gives an absolute lower limit on the $\left\langle \theta\right\rangle$ value for emitting synchrotron radiation. For $\left\langle \theta\right\rangle$ smaller than this critical angle only UHECRs are emitted. See Fig.~\ref{referee1}a and Fig.~\ref{cones}.

The timescale of the process is in general set by the density of particles around the BH, which is provided by the structure of the cavity and SN ejecta, see section~\ref{sec:4}.

We have already shown that during each such elementary process the BH experiences a very small fractional change of angular momentum
\begin{eqnarray}
 & |\Delta J|/J\approx (|\dot{J}|/J) \tau_{\rm ob}\approx 10^{-16},\nonumber \\  
 &|\Delta M|/M \approx (|\dot{M}|/M)\tau_{\rm ob} \approx 10^{-16}.
 \label{eq:j}
\end{eqnarray}

The electromagnetic energy of the first impulsive event given above  is a small fraction of total extractable rotational energy of the Kerr BH, see Eq.~(\ref{Eextr1}):
\begin{equation}
   \frac{\cal E_{\rm 1}}{E_{\rm ext}}\approx  10^{-16}.
\end{equation}
This clearly indicates that the rotational energy extraction from Kerr BH: 

1) Occurs  in ``discrete quantized steps''.

2) Temporally separated by $10^{-14}$--$10^{-10}$~s.

3) The luminosity of the GeV emission in GRB 130427A is not describable by a continuous function as traditionally assumed: it occurs in a ``discrete sequence of elementary quantized events'' \citep{2019arXiv190708066R}. 

There are two main conclusions which can be inferred from the theory of synchrotron radiation implemented in this section:

1.
Synchrotron radiation is not emitted isotropically and is angle dependent, the smaller the angle the higher the synchrotron photon energy, see Eq.~(\ref{maxgev}) and Fig.~\ref{cones}.

2. The  energy emitted in synchrotron radiation before reaching its asymptotic value is a function of injection angle $\theta$, see first line of Eq.~(\ref{Esyncsol}) and Fig.~\ref{referee1}b.

In order to compare and contrast the results based on this essential theoretical treatment with observation we need  to determine 1) the number of electrons for each selected injection angle $\theta$ and 2) to  verify that the  radiated synchrotron energy is compatible with the electrostatic energy  for each electron. 3) Taking to due account the role of the beaming angle which we have here derived, see Fig.~\ref{cones} {and extending the number of parameters by allowing the anisotropic distribution of electrons.}

\begin{figure*}
\centering
\gridline{\fig{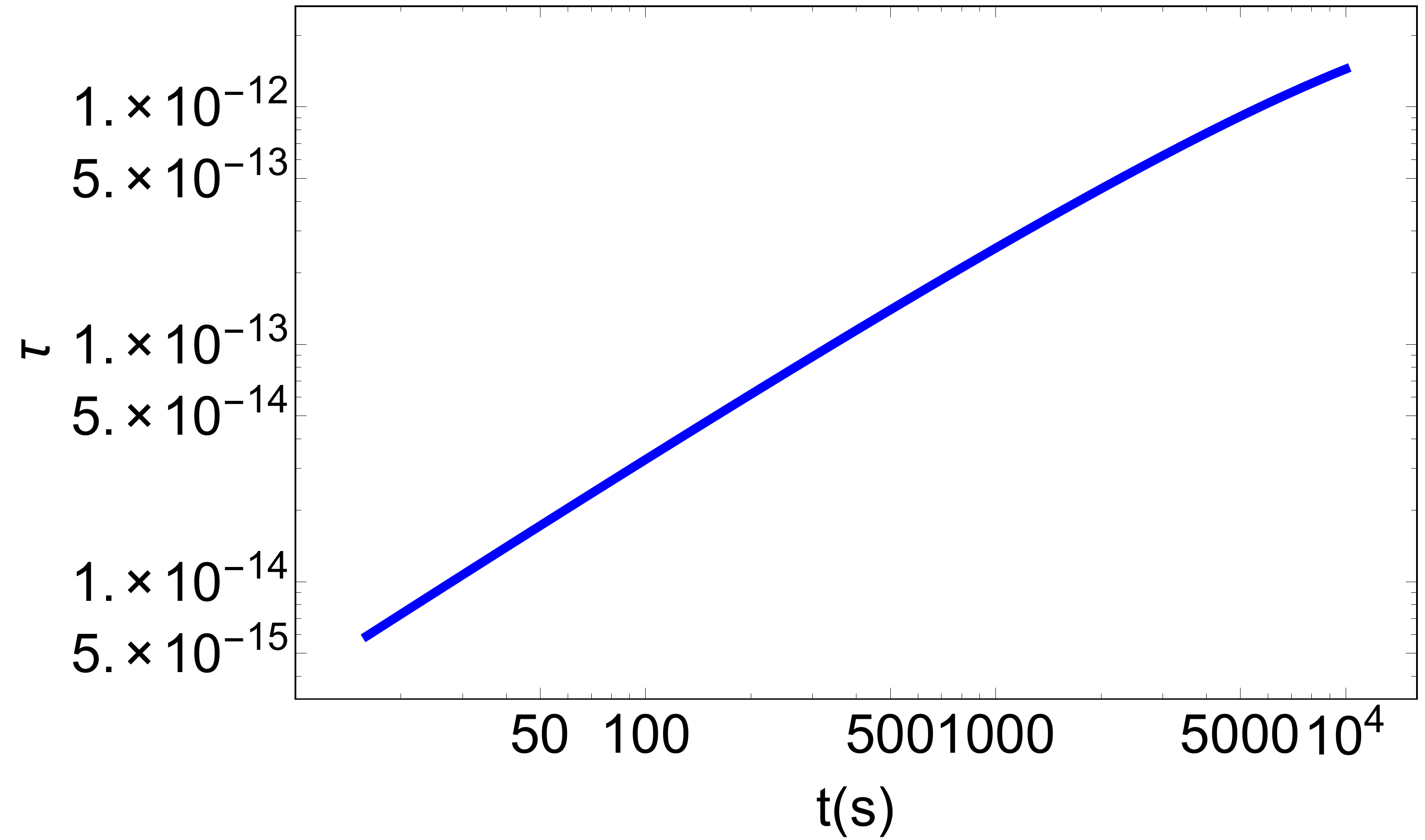}{0.48\textwidth}{(a)}
\fig{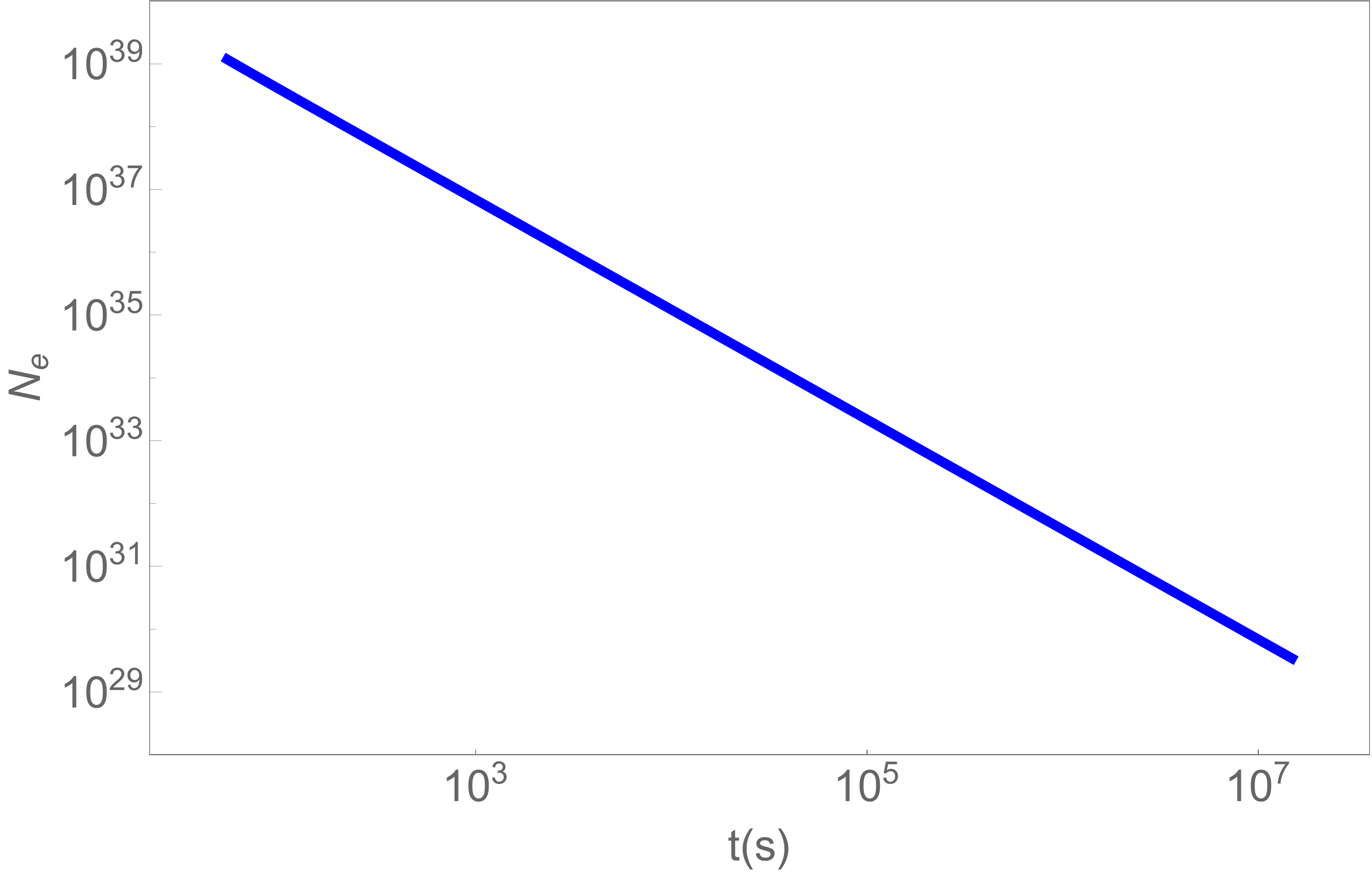}{0.45\textwidth}{(b)}}
\caption{a: The value of $\tau_{\rm ob}(t)=E_{r_+}^2 r_+^3/(2 L_{\rm GeV})$ calculated from the GeV luminosity data obtained from \textit{Fermi}-LAT together with the values of {the electric field and the horizon radius, $E_{r_+}$ and $r_+$,} obtained in each \textit{impulsive event}. This timescale increases linearly with the time $t$ (in seconds) as {$\tau_{\rm ob}\approx 3\times 10^{-16}~t^{1}$. b: Number of electrons available in each impulse to  fulfill the observed properties of the \textit{inner engine} of GRB 130427A.}}
\label{fig:tau}
\end{figure*}

\section{The repetition time of sequence of the discrete ``\textit{elementary impulsive events}''  }\label{sec:6}

We now finally study the sequence of iterative \textit{impulsive events} in which the system starts over, with a new value of the electric field set by the new values of the BH angular momentum and mass, $J = J_0-\Delta J$ and $M = M_0-\Delta M$, keeping the magnetic field value constant $B_0$. 

We infer from the  luminosity the evolution of the timescale $\tau(t)$  of the repetition time of the \textit{impulsive events} by requiring it to explain the GeV emission, i.e.:
\begin{equation}
 L_{\rm GeV}=\frac{{\cal E}}{\tau (t)},
\end{equation}
where ${\cal E}$ is the electrostatic energy {available for each impulsive event}. Therefore we obtain for the timescale
\begin{equation}\label{tauinner}
  \tau(t)=\frac{1}{2} \frac{E_{r_+}^2 r_+^3}{ L_{\rm GeV}},
\end{equation}
where $E_{r_+}$ is the electric field evaluated at the horizon, determined from the new values of $J$ and $M$ {for each elementary impulsive event consisted with Eq.~(\ref{eq:j})}. Fig.~\ref{fig:tau} a shows that $\tau_{\rm ob}$ is a  increasing  power-law function of time
\begin{equation}
 \tau_{\rm ob} \propto \frac{\alpha^2}{L_{\rm GeV}}\propto  t.
 \label{tauinner1}
\end{equation}

We identify the timescale $\tau_{\rm ob}$ with the repetition time of {each} impulsive event. The efficiency of the system diminishes with time, as shown by the increasing value of $\tau_{\rm ob}$ (see Fig.~\ref{fig:tau}a). This can be understood by the evolution of  the density of particles near the BH {which} decreases {in time} owing to the expansion of the SN remnant, making the iterative process become less efficient. As we have mentioned in the immediate vicinity of the BH a cavity is created of approximate radius $10^{11}$~cm and with very low density on the order of $10^{-14}$~g~cm$^{-3}$ \citep{2019ApJ...883..191R}. This implies an approximate number of $\sim 10^{47}$~ electrons inside the cavity. Then, the electrons of the cavity can power the iterative process only for a short time of $1$--$100$~s. We notice that at the beginning of the gamma-ray emission the required number of electrons per unit time for the explanation of the prompt and the GeV emission can be as large as $10^{46}$--$10^{54}$~s$^{-1}$. This confirms that this iterative process has to be sustained by the electrons of the remnant, at $r\gtrsim 10^{11}$~cm, which are brought from there into the region of low density and then into the BH. 

{ The synchrotron time scale, obtained from Eq.~(\ref{tcsynch}), has values between $1.6\times 10^{-14} $~s to $2 \times 10^{-5}$~s, see Fig.~\ref{referee1} (a). According to Eq.~(\ref{Esyncsol}) and Fig.~\ref{referee1} (b), the total energy available for the synchrotron radiation of one electron is a function of synchrotron time scale. }

{Therefore, when $t_c= 1.6 \times 10^{-14}$~s  the total energy available for each electron is $\epsilon_e = 1.02 \times 10^{9}$~eV which leads to the total number of electrons from Eq.~(\ref{enumber}), $N_e = {\cal E_{\rm 1}}/\epsilon_e \approx 6.6\times 10^{39}$. When $t_c=2 \times 10^{-5}$~s, the total energy available for each electron is $\epsilon_{e,max} = 1.65 \times 10^{18}$~eV and the total number of electrons is  $N_e = {\cal E_{\rm }}/\epsilon_{e,f} \approx 4\times 10^{30}$ which can be seen in Fig.~\ref{fig:tau} (b).}

It is worth to mention that the TeV synchrotron photons {in this picture} will start to be produced at $t_{\rm c}\sim 10^{-11}$~s see Fig.~\ref{referee1} (a), therefore, according to Eq.~(\ref{tauinner1}) the onset of TeV photons for GRB 130427A should be around $10^{5}$~s; see Fig.~\ref{fig:tau}. {This is clearly a zeroth-order approximation since considering the effect of angle-dependent distribution of electrons, this time could be shorter (J. A. Rueda, et al. 2019, in preparation). In any case, the feedback from the observations is needed in order to improve the model.}

\section{Conclusions}\label{sec:7}

{In this paper we confirm that the high energy GeV radiation observed by Fermi LAT originate from the rotational energy of a Kerr BH of mass $M=2.3 M_{\odot}$ and spin parameter of $\alpha=0.47$ immersed in an homogeneous magnetic field of $B_0 \sim  3.48 \times 10^{10}$~G{, and an ionized plasma of very low density $10^{-14}$~g~cm$^{-3}$ \citep{2019ApJ...883..191R}}. The radiation occurs following the formation of the BH, via synchrotron radiation emitted by ultra-relativistic electrons accelerating and radiating 
in a sequence of elementary impulses.}

{In the traditional approach see e.g. \citep{zhang_2018} there is a blast wave originating in the prompt radiation phase and propagating in a ultra relativistic jet into the ISM medium emitting synchrotron radiation   following the approach of \citet{1999ApJ...519L..17S}. There the kinetic energy of the blast wave is used  as the energy source. This process of emission occurs at large distances typically  of $10^{15}$~cm to $10^{16}$~cm and it is traditionally used to explain the GeV emission  observed by Fermi LAT as well as the  X-ray afterglow emission observed by SWIFT and  the radio emission observed by radio interferometers like Westerbork Synthesis Radio Telescope (WSRT).}
 
{In our approach {the} model is only a function of three parameters: the mass $M$ and spin $\alpha$ of the Kerr BH and the background magnetic test field which have been self  consistently derived in this article. They fulfill all the energetic and transparency requirements. The acceleration and synchrotron radiation process occurs within $10^5$~cm  from the horizon. The photon energy emitted by the synchrotron radiation process is a very strong function of the injection angle of the ultra relativistic electrons with respect to the polar axis.  The outcome is an emission over a large angle, up to  $\theta=\pi/3$: GeV at angle $\theta=\pi/8$ and TeV at $\theta=10^{-4}$  all the way up to UHECR at angle $\theta<10^{-11}$.  The particles accelerated by the electromagnetic field gyrate into the magnetic field $B_0$. The highly anisotropic distribution in the energy and in the spectra is a specific consequence of the current model.}

{A byproduct of our model has been to evidence for the first time that the high energy emission of GRB 130427A is not emitted continuously but in a repetitive sequence of discrete and quantized ``elementary impulsive events'' each of energy $10^{37}$~erg and with a repetition time of $\sim 10^{-14}$~s and slowly increasing with time. This implies a very long time of extraction of the BH rotational energy via this electromagnetic process each utilizing a fraction of $\sim 10^{-16}$ of the mass-rotational energy of the BH. This result was truly unexpected and it appears to be a general property both of GRBs and of much more massive Kerr BHs in Active Galactic Nuclei (AGN). The results obtained in this paper are leading to the concept of a ``Blackholic Quantum'' affecting our fundamental knowledge of physics and astrophysics \citep{2019arXiv190708066R}.}

{It is appropriate here to recall that within BdHN model the GeV and TeV emission observed by  Fermi-LAT and MAGIC detectors, originating from the Kerr BH, have a separate origin from the X-ray and radio observation of the afterglow. As recently demonstrated in five BdHN, GRB 130427A  GRB 160509A GRB 160625B, GRB 180728A and GRB 190114C \citep{2019arXiv190511339R},  the afterglow emission occurs due to the accretion process of the hypernova ejecta on the  $\nu$NS  spinning with a few millisecond period and the associated synchrotron emission.}

\acknowledgments
We thank the Referee for the detailed reports and precise questions which have motivated a more precise formulation of our paper. We are grateful to Prof. G.~V.~Vereshchagin and to S.~Campion for discussions in formulating the synchrotron radiation considerations and the related figures in the revised version. We would also like to thank Prof. Sang Pyo Kim  for usefull discussion about overcritical field in such a field. M.K. is supported by the Erasmus Mundus Joint Doctorate Program Grant N.2014--0707 from EACEA of the European Commission. N.S. acknowledges the support of the RA MES State Committee of Science, in the framework of the research project No. 18T-1C335.

\end{document}